\documentclass[floatfix,aps,prd,abstract,twocolumn,10pt,nofootinbib,preprintnumbers]{revtex4-1}

\usepackage{amsgen,amsmath,amstext,amsbsy,amsopn,amssymb}
\usepackage{graphicx}
\usepackage[usenames]{color}
\usepackage{epsfig}
\usepackage{multirow}
\usepackage{l  ongtable}

\newcommand{\mb}{\mathbf}

\newcommand{\Ddel}{\delta_{\rm D}   }

\newcommand{\MpcOh}{ \,  \mathrm{Mpc}  \, h^{-1} }
\newcommand{\hOMpc}{ \,  \mathrm{Mpc}^{-1}  \, h  }
\newcommand{\bP}{\mb{ \Psi} }
\newcommand{\comment}[1]{}

\begin{document}


\title{ Helmholtz Decomposition of the Lagrangian Displacement }

\author{Kwan Chuen Chan}\email{KwanChuen.Chan@unige.ch}
\affiliation{D\'epartement de Physique Th\'eorique and Center for Astroparticle Physics,
Universit\'e de Gen\`eve, 24 quai Ernest Ansermet, CH--1211 Gen\`eve 4,
Switzerland}

\date{\today}

\begin{abstract}
Lagrangian displacement field $\bP  $ is the central object in Lagrangian perturbation theory (LPT). LPT is very successful at high redshifts, but it performs poorly at low redshifts due to severe shell crossing.  To understand and quantify the effects of shell crossing, we extract $\bP $ from $N$-body simulation and decompose it into scalar and vector parts.  We find that at late time the power spectrum of the scalar part agrees with 1-loop results from LPT at large scales, while the power in small scales is much suppressed due to shell crossing.  At  $z=0$, the power spectrum of $\bP$ is 10\% lower than the 1-loop results at $k =0.1  \hOMpc $. Shell crossing also generates the vector contribution in $ \Psi  $, although its effect is subdominant in comparison with the power suppression in the scalar part. At $z=0$, the vector part contributes $10 \%  $ to the total power spectrum of  $\bP $ at $k = 1 \ \hOMpc $, while only 1\% is expected from the vector contribution in LPT.  We also examine the standard LPT recipes and some of its variants. In one of the variants,  we include a power suppression factor in the displacement potential to take into account the power suppression in small scales after shell crossing.  However, these simple phenomenological approaches are found to yield limited improvement compared to the standard LPT after the onset of shell crossing.  
\end{abstract}

\maketitle

\section{Introduction}

With the upcoming large-scale surveys, such as Euclid and LSST, a huge number of mock catalogs have to be generated to estimate the covariance matrix. Straight $N$-body simulations are too numerically expensive to be done in mass production, so many semi-analytic approaches such as PThalos \cite{ScoccimarroSheth2002,Maneraetal2013} PINOCCHIO \cite{Monacaetal2002,Monacoetal2013,Heisenbergetal2011}, and  COLAR \cite{Tsssevetal2013}, have been developed.  These methods rely on Lagrangian perturbation theory (LPT) to displace the particles at large scales.

In LPT, the fundamental object is the Lagrangian displacement field $\bP$, which displaces the particles from the initial position $\mb{q}  $ to its final Eulerian position $\mb{x} $
\begin{equation}
\label{eq:PsiDefinition}
\bP( \mb{q}, t  ) \equiv  \mb{x}( \mb{q}, t ) - \mb{q}.  
\end{equation} 

$\bP  $ can be computed using LPT. The first order LPT is the well-known Zel'dovich Approximation (ZA) \cite{Zeldovich1970} and it has been extended to higher orders \cite{Buchert1994,Catelan95,CatelanTheuns96,Bouchetetal1995,RampfBuchert12,Rampf2012}. The initial conditions for $N $-body simulations are often generated using ZA or second order LPT (2LPT) \cite{Scoccimarro98,CroccePeublasetal2006}. The validity of LPT in computing the power spectrum has been improved by resummation \cite{Matsubara08a}, and it can be easily extended to include redshift space distortion and local Lagrangian bias \cite{Matsubara08b,Matsubara11}.

 LPT is very successful at high redshifts but it yields poor results at late times due to severe shell crossing. Shell crossing occurs when particles from different Lagrangian patches meet to form caustics and multiple streams pass through the same Eulerian position. The standard perturbation theory and LPT are based on the single stream approximation \cite{PTreview}.  Before shell crossing, the system can be described by a velocity field. However, shell crossing generates the velocity dispersion tensor in small scales, which also sources the vorticity  \cite{PueblasScoccimarro2009}. In LPT, the velocity field is also specified by the position $\mb{x} $, so it is not valid after shell crossing. Indeed the Eulerian density obtained from LPT becomes singular after shell crossing \cite{ShandarinZeldovich1989}. After shell crossing, the particles keep on escaping from each other, resulting in low power in small scales.  For example, the ZA dark matter density power spectrum is even lower than the linear one at $z=0$. There are many attempts to extend the validity of LPT after shell crossing \cite{MelottPellmanetal1994, KitauraSreffen2012, Leclerceetal2013}.  One of the concrete models that takes shell crossing into account is the adhesion model \cite{Gurbatovetal1989}, in which a viscous term is added to the ZA model to stick the particles together after shell crossing, and the equation is transformed to Burgers' equation  \cite{Gurbatovetal1989,Vergassolaetal1994,Valageas2011,ShandarinZeldovich1989}. Note that the solution to the Burgers' equation is a velocity field. In practice, to get the position of the particles, one still needs to integrate the velocity field numerically  \cite{WeinbergGunn}. Even in the limit of zero viscosity, the geometrical construction is not so straightforward \cite{Gurbatovetal1989,Vergassolaetal1994}.

 There are few studies on $\bP  $ directly, if any.  In this paper, we shall extract  $\bP $  from $N$-body simulation directly, and this will enable us to probe  $\bP $ even after shell crossing.  To study $ \bP $ directly is interesting because $ \bP $ is the fundamental object in LPT and  there are  few analytical tools available to study $\bP $ after shell crossing.  The goal of this paper is to better  understand the physics of shell crossing on  $\bP$ by examining  $\bP $  obtained from simulation. This may potentially lead to better modeling of LPT at late time.  We will also examine some modifications of LPT. In particular, we attempt to incorporate the power suppression in small scales due to shell crossing with an effective potential.  It turns out that, as shell crossing is a highly nonlinear process, this phenomenological approach is of limited success.  In LPT, $\bP $ is often taken to be potential.  Another question that we want to address in this paper is whether the potential assumption is still valid at low redshifts. In particular, we will quantify how important the vector part of $ \bP $ is at late time, when LPT is known to break down.  We shall decompose the numerical $\bP $ into scalar and vector parts.   Very often in the studies related to LPT, when compared with simulation, only the density power spectrum is considered. This is justifiable as the density field is the final observable. However, as $\bP $ plays a central role in LPT, we believe that studying it in its own right is worthwhile.

The paper is organized as follows. We will describe the decomposition method in Sec.~\ref{sec:HelmholtzPsiGeneral}. LPT is reviewed, and the loop  corrections to power spectrum of $\bP  $ are  written down in Sec.~\ref{sec:LPTreview}. The numerical results for the decomposition of $\bP $ are presented in Sec.~\ref{sec:NumericalResults}.  We will show the scalar and vector power spectrum of $\bP  $ in details in Sec.~\ref{sec:PkPsiNumerical}.  In Sec.~\ref{sec:ModificationLPT}, we examine LPT and a couple of variants of LPT using density power spectrum. In particular, we  include a suppression factor in the displacement potential to modify LPT.   We explore the scatter plot of $\nabla \cdot \Psi  $ in Sec.~\ref{sec:ScatterDivPsi}.  We conclude in Sec.~\ref{sec:Conclusions}. The general structure of the power spectrum of $\bP $ is written down in Appendix \ref{sec:GeneralPsiPk}.  In Appendix \ref{sec:TestCases}, we test the decomposition algorithm with some test cases.

\section{Helmholtz decomposition of $ \bP $ and its power spectra}

\subsection{Helmholtz decomposition of $ \bP $ }
\label{sec:HelmholtzPsiGeneral}

Any smooth vector field $\bP   $ can be decomposed into the form \footnote{The uniqueness of the decomposition generally depends on the boundary conditions. If it is not unique, the difference is due to a harmonic part which is both divergence-less and curl-free, or it can be written in terms of a potential, which is harmonic. In electromagnetism, if we require that the field vanishes at infinity, then because harmonic function cannot have local extremum, it must vanish. Here, we impose the periodic boundary condition in simulation, the only smooth function that satisfies the periodic boundary condition without local extremum in each dimension is a constant function.  If we further require the field to have zero mean, then it must vanish everywhere.    }
\begin{equation}
\bP = \nabla  \Phi + \nabla \times \mb{A} 
\end{equation}
where $\Phi$ is the scalar potential and  $\mb{A}$ is the vector potential. We stress that the derivatives are with respect to the Lagrangian coordinates. This kind of decomposition has been widely used in physics, for example, in the decomposition of the electric field in electromagnetism \cite{Jackson}, and the cosmological perturbation theory \cite{Bertschinger1995}. Recently, it has been applied to redshift space distortion as different components correspond to different physical origins \cite{ Zhangetal2013,Zhengetal2013}.  The scalar and vector potentials can be solved through the Poisson equations 
\begin{eqnarray}
\label{eq:Poisson_Phi}
\nabla^2 \Phi &= &\nabla \cdot \bP  , \\
\label{eq:Poisson_A}
\nabla^2 \mb{A} &=& - \nabla  \times  \bP  . 
\end{eqnarray}

In Fourier space, the helicity basis is convenient for decomposing of $ \bP  $ into scalar and vector parts. The helicity basis vectors are defined as 
\begin{eqnarray}
\label{eq:Helicity0}
\hat{ \mb{k}}_0 & = & \hat{ \mb{k} },   \\
\label{eq:HelicityPlus}
\hat{ \mb{k}}_{+} &=&  \frac{1 }{ \sqrt{2} }( \hat{\mb{k}}_{\theta } + i \hat{\mb{k}}_{\phi} ),  \\
\label{eq:HelicityMinus}
\hat{ \mb{k}}_{-} &=&  \frac{1 }{ \sqrt{2} }( \hat{\mb{k}}_{\theta } - i \hat{\mb{k}}_{\phi} )  ,
\end{eqnarray}
where $ \hat{ \mb{k} }$, $  \hat{\mb{k}}_{\theta }  $ and  $\hat{\mb{k}}_{\phi} $ are the basis vectors in spherical coordinates. Then the scalar part is given by the helicity-0 mode,  $\hat{ \mb{k}}_0$ component, and the vector part is decomposed into the  helicity-$\pm  $ modes, the  $\hat{ \mb{k}}_{+}$ and $\hat{ \mb{k}}_{-}$ components. We shall make use of this basis in measuring the power spectrum.    We use the terminology, scalar and vector decompositions and longitudinal and transverse parts, potential and curl parts, and helicity-0 and helicity-$ \pm  $ interchangeably in this paper. 

We stress that in standard LPT, the displacement field is almost fully potential. At late time LPT is known to break down due to severe shell crossing. Thus the generation of the vector part in $\bP   $ can help understand shell crossing and  shred light on the break down of LPT at late time.

\subsection{$ \bP $ from Lagrangian Perturbation Theory}
\label{sec:LPTreview}

We review LPT in this section. To facilitate the comparison with numerical power spectrum of $\bP$, we shall write down the loop  corrections to the power spectrum of $\bP $ from LPT. We will also describe the recipes to generate LPT catalogs numerically. We emphasize that the review of LPT here serves as a check and comparison with the numerical results shown later on; in this paper, we are more interested in exploring the effects that are not captured by the LPT discussed here. 

\subsubsection{Power spectrum of $\bP   $ from LPT   }

In Appendix \ref{sec:GeneralPsiPk}, we show the general structure of the power spectrum of the scalar and vector components. Here we will write down the 1-loop power spectrum of $\bP$ from LPT. 

This displacement field $\bP  $ can be expanded in terms of the linear dark matter density contrast  in LPT.  Up to third order, it is given by 
\begin{equation}
\label{eq:PsiExpansion}
\bP = \bP^{ (1) } + \bP^{ (2) } + \bP^{ (3a) } + \bP^{ (3b) } + \bP^{ (3c) } ,
\end{equation} 
where
\begin{eqnarray}
\bP^{( n )}  (\mb{k}, t)  & = &  i D^n(t)  \int  d^3 p_1 \dots d^3 p_n  \Ddel ( \mb{k} - \mb{p}_{1 \dots n } ) \\
&\times &  \mb{L}^{ (n) } ( \mb{p}_1,  \dots , \mb{p}_n ) \delta_0 ( \mb{p}_1 ) \dots \delta_0 ( \mb{p}_n),   \nonumber
\end{eqnarray}
where $D$ is the linear growth factor and $ \mb{p}_{1 \dots n } $ denotes $\mb{p}_1 + \dots + \mb{p}_n  $,  $\delta_0 $ is the initial linear dark matter density contrast, and   $  \Ddel $ is the Dirac delta function.  The Lagrangian displacement kernels are given by \cite{Catelan95, CatelanTheuns96, Matsubara08a, RampfBuchert12}
\begin{eqnarray}
\label{eq:1LPTZA}
\mb{L}^{(1)} ( \mb{p}_1 )  &=&   \frac{ \mb{p}_1  } {  p_1^2 }  ,\\
\label{eq:2LPT}
\mb{L}^{(2)} ( \mb{p}_1,\mb{p}_2)  &=&    \frac{ 3  }{ 14 }  \frac{  \mb{p}_{12}   } {  p_{12}^2 }  \Big[ 1 - \frac{ ( \mb{p}_1 \cdot  \mb{p}_2 )^2 }    { p_1^2 p_2^2  }   \Big] , \\
\mb{L}^{(3a)}_{\rm a} ( \mb{p}_1, \mb{p}_2, \mb{p}_3 )  & = &  - \frac{ 1 }{ 18 } \frac{ \mb{p}_{123}  } {  p_{123}^2 }  \Big[ 1 - 3 \frac{ ( \mb{p}_1 \cdot   \mb{p}_2 )^2   }{ p_1^2  p_2^2  }   \\
& +&  2 \frac{ (\mb{p}_1 \cdot \mb{p}_2)(  \mb{p}_2 \cdot \mb{p}_3)(   \mb{p}_3 \cdot \mb{p}_1 ) } { p_1^2  p_2^2 p_3^2  }      \Big]     ,   \nonumber   \\
\mb{L}_{\rm a}^{(3b) } ( \mb{p}_1,\mb{p}_2,\mb{p}_3 )  &=&    \frac{ 5 }{ 42 }   \frac{ \mb{p}_{123}  } {  p_{123}^2 }   \Big[  1 - \frac{ ( \mb{p}_1 \cdot  \mb{p}_2 )^2 }{ p_1^2 p_2^2 }   \Big]    \\
& \times  &    \Big[  1 -  \Big(  \frac{  \mb{p}_{12}   \cdot \mb{p}_3  }{ p_{12}  p_3  }      \Big)^2    \Big]    ,       \nonumber    \\
\mb{L}^{(3c)}_{\rm a} ( \mb{p}_1, \mb{p}_2, \mb{p}_3 )  & = &    \frac{ 1  }{ 14  } \frac{ \mb{p}_1 \cdot  \mb{p}_{23}  }{ p_1^2   p_{23}^2  p_{123}^2 }    \Big[    1 - \frac{ ( \mb{p}_2 \cdot  \mb{p}_3 )^2 }{ p_2^2 p_3^2 }   \Big]     \\ 
& \times &   [   \mb{p}_1 ( \mb{p}_{123} \cdot  \mb{ p }_{23}  )   -     \mb{p}_{23} ( \mb{p}_{123} \cdot  \mb{ p }_{1}  )  ] . \nonumber 
\end{eqnarray}
The kernels $\mb{L}^{(3)}_{\rm a} $ are asymmetric with respect to the arguments, and we will symmetrize them as 
\begin{equation}
\mb{L}^{(3)} (\mb{p}_1,  \mb{p}_2 , \mb{p}_3  ) = \frac{ 1  }{ 3 } [  \mb{L}^{(3) }_{\rm a} (  \mb{p}_1,  \mb{p}_2 , \mb{p}_3   )  +   2 \, \rm{cyc.}     ] .  
\end{equation}
The first order kernel Eq.~\ref{eq:1LPTZA} corresponds to the 1LPT, \textit{i.e.}~the ZA \cite{Zeldovich1970}, and Eq.~\ref{eq:2LPT} is the 2LPT. Except $\mb{L}^{(3c)}$, all the other kernels are proportional to $\mb{p}_{1,\dots n} $, where $n$ is the order, and so they are potential. Note that in LPT, it is still a potential flow in Eulerian space, and the appearance of the curl part kernel $\mb{L}^{(3c)}$ is due to the coordinate transformation from the Lagrangian space to the Eulerian space \cite{Catelan95}.

The power spectrum of $\bP$  is defined as 
\begin{eqnarray}
\langle \bP_i (\mb{k}_1 ) \bP_j (\mb{k}_2) \rangle = P_{ij}(k_1) \Ddel( \mb{k}_{12} ).
\end{eqnarray}
Using the expansion of $\bP$, Eq.~\ref{eq:PsiExpansion}, we can compute the power spectrum. Up to 1-loop, they are given by 
\begin{eqnarray}
\label{eq:PPsi11}
P_{ij}^{11}(k) &=&  D^2  \mb{L}_i^{(1)}( \mb{k} )  \mb{L}_j^{(1)}( \mb{k} )  P_0(k)  ,\\
\label{eq:PPsi22}
P_{ij}^{22}(k) &=& 2 D^4 \int d^3 q  \mb{L}_i^{(2)}( \mb{q}, \mb{k} - \mb{q} )   \\
& \times &    \mb{L}_j^{(2)}( \mb{q}, \mb{k} - \mb{q} )  P_0 (q) P_0(|\mb{k} - \mb{q} |) ,  \nonumber  \\ 
\label{eq:PPsi13}
P_{ij}^{13}(k) &=&  6  D^4 P_0(k) \mb{L}_i^{(1)}(\mb{k}) \int d^3 q   \\
 &\times &  \mb{L}_j^{(3)} ( \mb{k} , - \mb{q} , \mb{q} ) P_0(q) ,  \nonumber 
\end{eqnarray}
where $P_0  $ is the initial power spectrum.  The integral in Eq.~\ref{eq:PPsi13} can be further simplified.  For the longitudinal part, we have
\begin{eqnarray}
&&  \int d^3 q   \mb{L}_j^{(3L)} ( \mb{k} , - \mb{q} , \mb{q} ) P_0(q)  = \frac{5 \pi }{3024  } \frac{ \mb{k} }{ k^5 }  \int  dq   \\  
& \times &  \frac{ P_0(q) }{ q^3 }  \Big[ -12 k^7 q + 44 k^5 q^3 + 44 k^3 q^5  \nonumber  \\
& -& 12 k q^7  +  3 (k^2 - q^2 )^4 \ln  \frac{(k+q)^2 }{(k-q)^2} \Big] . \nonumber   
\end{eqnarray}
For the transverse part, the integral  is given by
\begin{eqnarray}
&& \int d^3 q  P_0( q )   \frac{ \mb{k} \cdot \mb{q}   }{ 21 k^2 q^2 | \mb{k} + \mb{q } |^2 }   \\
& \times &  \Big(  1 - \frac{ ( \mb{q} \cdot \mb{k})^2 }{q^2 k^2 }      \Big)   \mb{k} \times ( \mb{q} \times \mb{k}    ),   \nonumber
\end{eqnarray}
which vanishes upon integration.  In fact,  this follows from the argument given in Appendix ~\ref{sec:GeneralPsiPk} that the cross power spectrum  between the scalar and vector part vanishes. Therefore, the lowest order vector contribution to the power spectrum of $ \bP$ arises from the auto power spectrum of $\bP^{ 3 c}  $, and it is a 2-loop contribution.  The lowest order vector  contribution reads 
\begin{eqnarray}
\label{eq:PPsi33v}
P^{33 \rm v}_{ij} (k) &= &6 D^6  \int d^3 q_1  \int d^3 q_2   P_0 (q_1)  P_0 (q_2)  P_0 ( |  \mb{k} -  \mb{ q}_{12} | )   \nonumber  \\
 & \times &   \mb{L}^{3 c}_i (\mb{q}_1, \mb{q}_2,    \mb{k} -  \mb{ q}_{12}  )  \mb{ L}^{3c}_j (\mb{q}_1, \mb{q}_2,    \mb{k} -  \mb{ q}_{12}  ) .  
\end{eqnarray}

\begin{figure*}[!htb]
\centering
\includegraphics[ width=\linewidth]{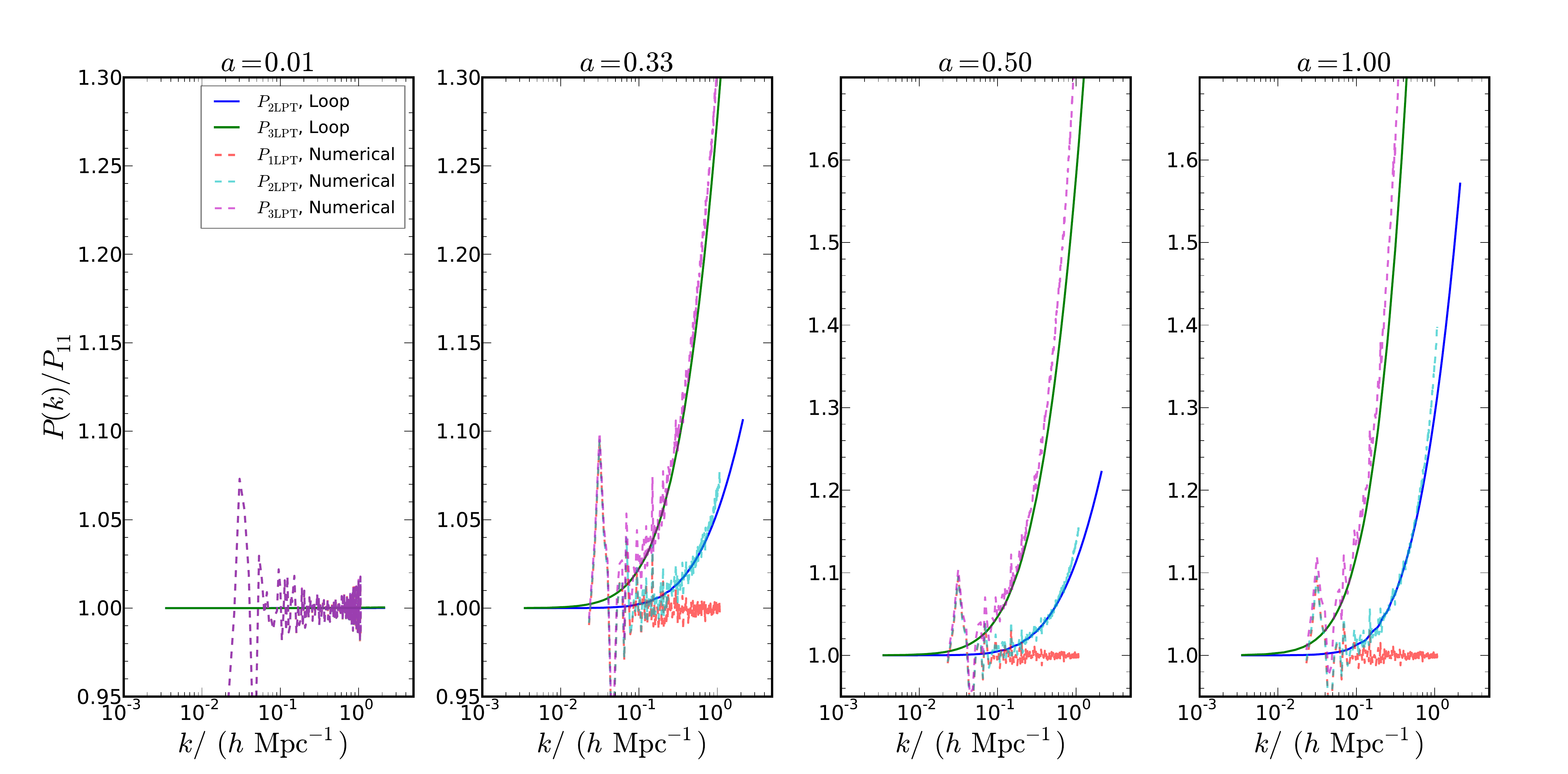}
\caption{ The 1-loop corrections to the power spectrum of $\bP  $ at different scale factors $a=0.01, \, 0.33, \, 0.5$  and 1 (from left to right). The 2LPT power spectrum (solid, blue) include $P^{22}$, while the 3LPT power spectrum (solid, green) further includes $P^{13}$. Numerical power spectrum of $\Psi  $ that generated by 1LPT (dashed, red), 2LPT (dashed, cyan) and 3LPT (dashed, violet) are also shown.  They are normalized with respect to the ZA power spectrum $P^{11}$.     }
\label{fig:Pkratio_1LoopNumerical}
\end{figure*}

In Fig.~\ref{fig:Pkratio_1LoopNumerical}, we show the 1-loop power spectrum of $\bP$, normalized by the tree level ZA power spectrum. As $\bP  $ is a vector, there are numerous ways to present its power spectrum. Here we show
\begin{equation}
P(k) = \sum_{i}  P_{ii} (k) ,
\end{equation}
because it is rotationally invariant and coordinate-independent.

At high redshift the loop correction terms are negligible, they however become important at low redshifts.  In particular we note that the contribution of $P^{13} $, which arises from 3LPT,  is much more significant than $P^{22}$, which appears in 2LPT. At $z=0$, $P^{13} $ is of 10\% of the ZA power spectrum at $k=0.1 \hOMpc $, while $P^{22}$ is only 1\% at this scale.  As we will see later on, including $P^{13} $, the agreement with the numerical $\bP $ is much improved at the weakly nonlinear regime, although it quickly causes more rapid deviation from the numerical results due to the onset of shell crossing in the weakly nonlinear regime.

\begin{figure}[!htb]
\centering
\includegraphics[ width=\linewidth]{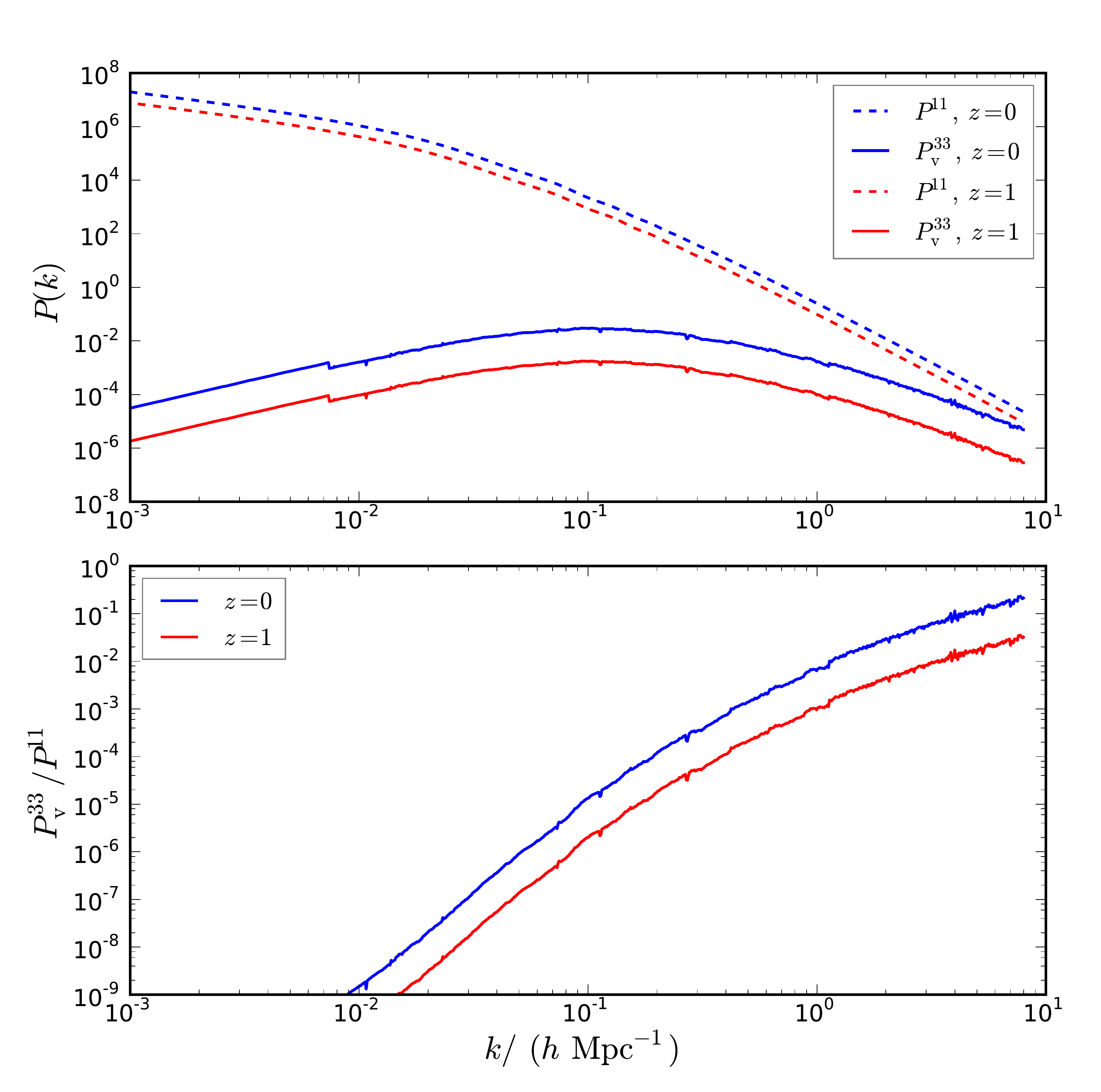}
\caption{ In the upper panel, the lowest order vector contribution  to the power spectrum of $\bP $,  $P^{33 \rm v}$ (solid) and the ZA power spectrum (dashed) are plotted. In the lower panel, the ratio  $P^{33 \rm v}  / P^{11}  $ is shown.   Two redshifts are shown: $z=0$ (blue) and $z=1$ (red). }
\label{fig:PkPsiV_2Loop}
\end{figure}

We plot the vector power spectrum  $P^{33 \rm v} $  and $P^{11}$ for comparison  in Fig.~\ref{fig:PkPsiV_2Loop}. Although the two-loop contribution grows much faster than the ZA power spectrum. At $z=0 $, $ P^{33 \rm v}  $ is only 1\% of the magnitude of  the ZA power spectrum at $k = 1 \, \hOMpc  $.  Thus the vector contribution from LPT to the power spectrum  of $\bP$ is small. However, we shall see  in Sec.~\ref{sec:NumericalResults} that at late time a much larger amount of the curl part is generated in small scales due to shell crossing. This non-perturbative effect is not captured by LPT.

\subsubsection{Generating Lagrangian displacement in simulations }
We will also generate the dark matter density field using LPT. Here we briefly review the procedures to generate the displacement field using LPT \cite{ Bouchetetal1995, Buchertetal1994, Scoccimarro98}.  Up to second order, the displacement field in LPT is potential. In third order, it acquires a curl part. However, we see in the previous section that it does not contribute to the power spectrum of $\bP $ at the 1-loop order. Thus we shall neglect the curl part, and the 3LPT displacement can be written in terms of the displacement potentials as 
\begin{equation}
\mb{ \bP }_{\rm 3LPT} = \nabla ( D_1 \phi^{(1)}  +  D_2 \phi^{(2)}  +  D_{\rm 3a} \phi^{(3 \rm a)}  +  D_{\rm 3b} \phi^{(\rm 3 b)}  ). 
\end{equation}
The LPT growth factors can be written in terms of the linear growth factor $D$ as
\begin{eqnarray}
D_1       &=& - D ,                 \\
D_2       &=& - \frac{3}{7} D^2,    \\
D_{\rm 3a} &=& - \frac{1}{3} D^3,     \\
D_{\rm 3b} &=& - \frac{10}{21} D^3.   
\end{eqnarray}
The displacement potentials are obtained by solving the following Poisson equations:
\begin{eqnarray}
\nabla^2 \phi^{(1)} & =&  \delta_0 ,  \\
\nabla^2 \phi^{(2)} & =& -\frac{1}{2} \mathcal{G}_2 ( \phi^{(1)} , \phi^{(1)} ) ,    \\
\nabla^2 \phi^{(3a)} & =& \det ( \nabla_{ij}  \phi^{(1)} ),    \\
\nabla^2 \phi^{(3b)} & =& -\frac{1}{2} \mathcal{G}_2 ( \phi^{(1)} , \phi^{(2)} ) , 
\end{eqnarray}
where $\mathcal{G }_2 (\phi^{(a)}, \phi^{(b)} ) $ denotes
\begin{equation}
\mathcal{G }_2 (\phi^{(a)}, \phi^{(b)} ) \equiv \sum_{i,j} \big(\nabla_{ij}\phi^{(a)} \nabla_{ij}\phi^{(b)}   \big)    - \nabla^2 \phi^{(a)} \nabla^2 \phi^{(b)} .
\end{equation}

In Fig.~\ref{fig:Pkratio_1LoopNumerical}, we also show the power spectrum of $\bP  $ obtained using the displacement field generated using 1LPT, 2LPT, and 3LPT, and they agree with the ZA and the loop corrections pretty well. 

We would like to comment that in the power spectrum from 3LPT catalogs, in addition to the 1-loop contributions, there is also the 2-loop scalar contribution $P^{33 \rm s  } $.  The good agreement between the 1-loop calculations and the results from 3LPT catalogs imply that the effects of  $P^{33 \rm s  } $ are negligible. This is in stark contrast to the standard perturbation theory, in which the individual contributions of the higher order loop terms give even more sizeable contribution than the lower order ones, although the total contributions are small due to large cancellations among the individual terms.

\section{Numerical results} 
\label{sec:NumericalResults}

In $N$-body simulation, since we know both the initial position $\mb{q}$ and the final position $\mb{x}$ of the particles,  we can easily extract $\bP $ using Eq.~\ref{eq:PsiDefinition}.   After getting  $\bP $, we can obtain the scalar and vector potentials by solving Eq.~\ref{eq:Poisson_Phi} and \ref{eq:Poisson_A} respectively. To compute the source $\nabla \cdot \bP  $ and $\nabla \times \bP $, we can either compute them using finite difference (FD) method in real space or using spectral derivative by means of Fast Fourier Transform (FFT) in Fourier space. In Appendix \ref{sec:TestCases}, we test the FD and FFT methods with some test cases, and we find that the FFT method performs better than the FD. Thus we shall use FFT method throughout this paper.  With the scalar and vector potentials, we can obtain the scalar and vector parts of $ \bP $.   Again we take the derivatives with the FFT method.  Another way to obtain the vector part of the field is simply to subtract the scalar part from the input field. We shall use both methods as crosschecks, and abbreviate the one obtained from vector potential as Vector and the one obtained by subtracting the scalar part from the input field as Input $-$ Scalar.  We shall see that both methods yield very similar results.

In the literature there have been measurements of scalar and vector part of the velocity field \cite{BernardeauWeygaert1996,WeygaertSchaap, PueblasScoccimarro2009,Zhengetal2013}. A major difficulty in velocity measurement is that it is sampled by discrete point particles. If the velocity field is obtained by interpolating the velocity of the particles to a grid, one would get a mass-weighted field rather than a volume-weighted one, \textit{i.e.}, one obtains momentum instead of velocity.  In the void region, the velocity is not necessarily small although there are few particles available for interpolation.  Various methods have been developed to cope with this problem, such as the Delaunay tessellation method \cite{BernardeauWeygaert1996,WeygaertSchaap, PueblasScoccimarro2009}.  However, for the measurement of $\bP$, since it is defined at all grid points, we do not have this sparse sampling problem.

As mentioned in Sec.~\ref{sec:HelmholtzPsiGeneral}, using the basis vectors Eq.~\ref{eq:Helicity0}--\ref{eq:HelicityMinus}, the fields are decomposed into the scalar (helicity-0 component) and the vector (helicity-$+$ and  helicity-$-$ components) automatically.  We will also use this method as a crosscheck.

Before presenting the numerical results, we shall first outline the details of the simulation used in this paper. In the simulation, there are $1024^3$ particles. Two box sizes are used, 1500 $\MpcOh $ and 250 $\MpcOh $. One realization for the 1500 $\MpcOh  $ box and three realizations for the 250 $\MpcOh  $ one. The cosmology is a  flat $\Lambda$CDM model, with the WMAP 7 cosmology parameters adopted \cite{WMAP7}, \textit{i.e.}, $\Omega_{\rm m} = 0.272$, $\Omega_{\Lambda}=0.728$, $\Omega_{\rm b} = 0.0455$, and  $\sigma_8=0.81$. Thus for the large box, each particle carries a mass of $2.37 \times 10^{11} \, M_{\odot} h^{-1} $ and  $1.10 \times 10^{9} \, M_{\odot} h^{-1} $ for the small box. The large  box enables us to probe the large scale mode.  On the other hand, as shell crossing is a small scale phenomenon, the small box simulation with better mass and spatial resolution will enable us to capture its effect more accurately.   The initial condition is Gaussian with spectral index being 0.967. The transfer function is output from CAMB \cite{CAMB} at redshift 99. The initial particle displacements are implemented using 2LPT \cite{CroccePeublasetal2006}. The simulation is done using Gadget2 \cite{Gadget2}. See \cite{Biagetietal2013} for more details.

\subsection{Numerical helicity power spectrum of $ \bP $ }
\label{sec:PkPsiNumerical}
\begin{figure*}[!htb]
\centering
\includegraphics[ width=\linewidth]{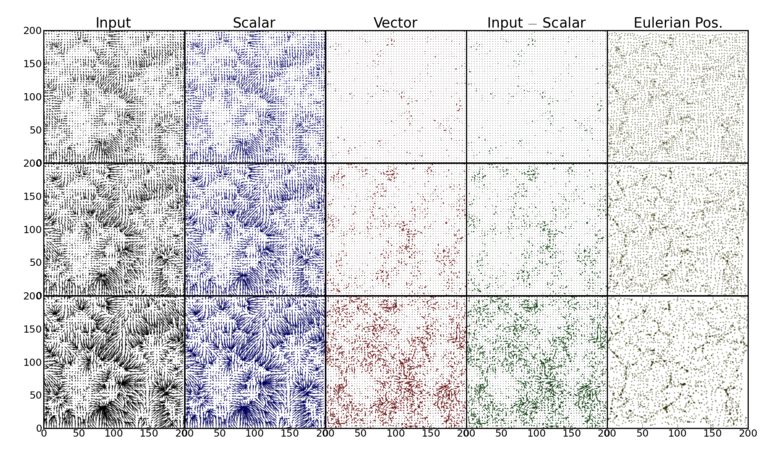}
\caption{ Sections of the vector fields. The fields are obtained from the 250 $\MpcOh $ box simulation. In each section, we show the projection of $\bP $ onto the $x-y$ plane. The foot of the arrow locates at the initial position $\mb{q}  $. The size of each section is 200 by 200 $(\MpcOh )^2$. The columns correspond to the original $\bP$ measured from numerical simulation, the scalar component of $\bP$, the vector component obtained by solving for the vector potential,  the vector field obtained by subtracting the scalar part from the original field and the Eulerian position of the particles (from left to right). Different rows are for $z=2$, 1 and 0 respectively (from top to bottom). The displacement fields are to the scale for Input and Scalar, but blown up by a factor of 5 for Vector and Input - Scalar.        }
\label{fig:vec_field}
\end{figure*}

We show in Fig.~\ref{fig:vec_field} the sections of the vector fields projected to the $x-y$ plane for the original $\bP $, its scalar component, the vector components obtained by solving the Poisson equation (Vector) and by subtracting the scalar components from the original field (Input $-$ Scalar). We also show the Eulerian positions of the particles.  First, the original input field is almost visually identical to its scalar component at all redshifts shown. The vector component is much smaller, and do not  have large-scale coherent component. We  note that the vector component fluctuates in signs at small scales, this qualitatively agrees with \cite{PichonBernardeau,PueblasScoccimarro2009}.   The vector components obtained in two different methods   result in  almost identical field pattern. In Appendix \ref{sec:TestCases}, we also find that these methods give very similar reconstruction results. By comparing the Eulerian positions of the particles with the plot of the vector part of $\bP$, it is clear that the vector parts are generated in the high density region where caustics form.  

\begin{figure*}[!htb]
\centering
\includegraphics[ width=\linewidth]{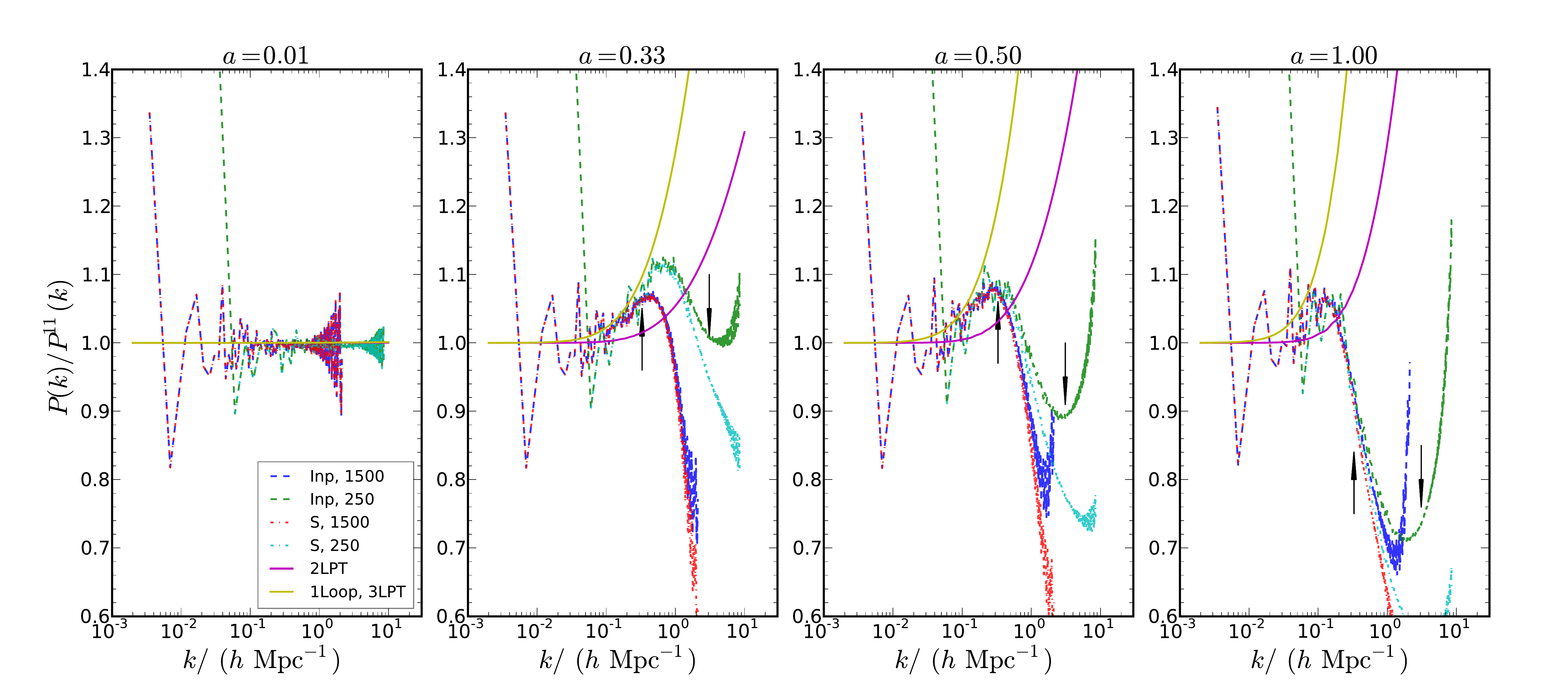}
\caption{ The power spectrum of $\bP $ from simulation, its scalar part, the 2LPT (solid, violet) and 3LPT (solid, yellow) results. The power spectrum $\bP $ from simulation (Inp) from two box sizes are shown, 1500 (dashed, blue) and 250 $\MpcOh $ (dashed, green). The scalar component (S) of $\bP $  from  the 1500 $\MpcOh  $ (dotted-dashed, red) and 250 $\MpcOh  $ (dotted-dashed, cyan) boxes are also shown.  The 250 $\MpcOh  $ box is averaged over three realizations, but the error bar is not shown for clarity.  The subplots from left to right are for $a=0.01$, 0.33, 0.5 and 1.  At low redshifts, around  $k\sim 1 \hOMpc   $, the 250 $\MpcOh  $ box, which has better resolution,  yields about 10\% higher power than the larger box. For each set of simulations, an arrow is added to suggest the scale above which there could be numerical artifacts.  }
\label{fig:PkPsi_SRatio}
\end{figure*}

We now turn to the power spectrum to study the decomposition more quantitatively.  In Fig.~\ref{fig:PkPsi_SRatio}, we compare power spectrum of $\bP  $ of the original field,  the power spectrum of its scalar part and, also the 2LPT and 3LPT loop power spectrum. At $z=99$, the initial conditions are set by 2LPT, which is completely potential, and indeed, the vector component is consistent with zero.   

At large scales, the results from the two boxes agree, however, at low redshifts, the small box yields higher power than the large one in small scales. At $k \sim 1 \MpcOh $, the smaller box results gives more than 10\% higher power than the large one. The smaller box with better mass and spatial resolution, it  measures  the effects of shell  crossing more accurately. Thus we shall trust the small box results in the large $k$ regime. We add an arrow to indicate the scale below which the large box results agree with the small box ones, and hence it is reliable. This scale is about 0.3 $\hOMpc $. For the smaller box, the increase in power in small scales is due to aliasing. We also add an arrow as a rough guide to indicate the scale, above which aliasing could be significant.

  The input field and its scalar component have the same power spectrum at large scales and deviation between them occurs only for large $k$ at low redshift.  At redshift 0, we find that the original field have higher power by 10\% at $k\sim 1 \, \hOMpc   $ than the scalar component.  Thus the scalar component is still the  dominant contribution at large scales even after shell crossing.  It is interesting to note that although the overall magnitude of the power spectrum of the original field and its scalar mode differs at this scale for the two box sizes, the ratio between the  original field and its scalar mode agrees quite well.

 At low redshifts, 3LPT gives higher power than both ZA and 2LPT  and it agrees with the numerical power spectrum better at the weakly nonlinear regime.  However, 3LPT keeps on shooting up, while the numerical power spectrum turns over due to shell crossing. In Fig.~\ref{fig:PkPsi_SRatio}, we first see a bump and then a sharp drop in power indicates that the scale that the nonlinear higher order corrections becomes important is larger than the shell crossing scale.  We also note that the turn-over scale decreases with time. At $z=0 $, the scale is around $0.1 \, \hOMpc   $. As the shell crossing scale increases, the effects of higher order LPT corrections only cause more rapid deviation from numerical results in the weakly nonlinear regime.

\begin{figure*}[!htb]
\centering
\includegraphics[ width=\linewidth]{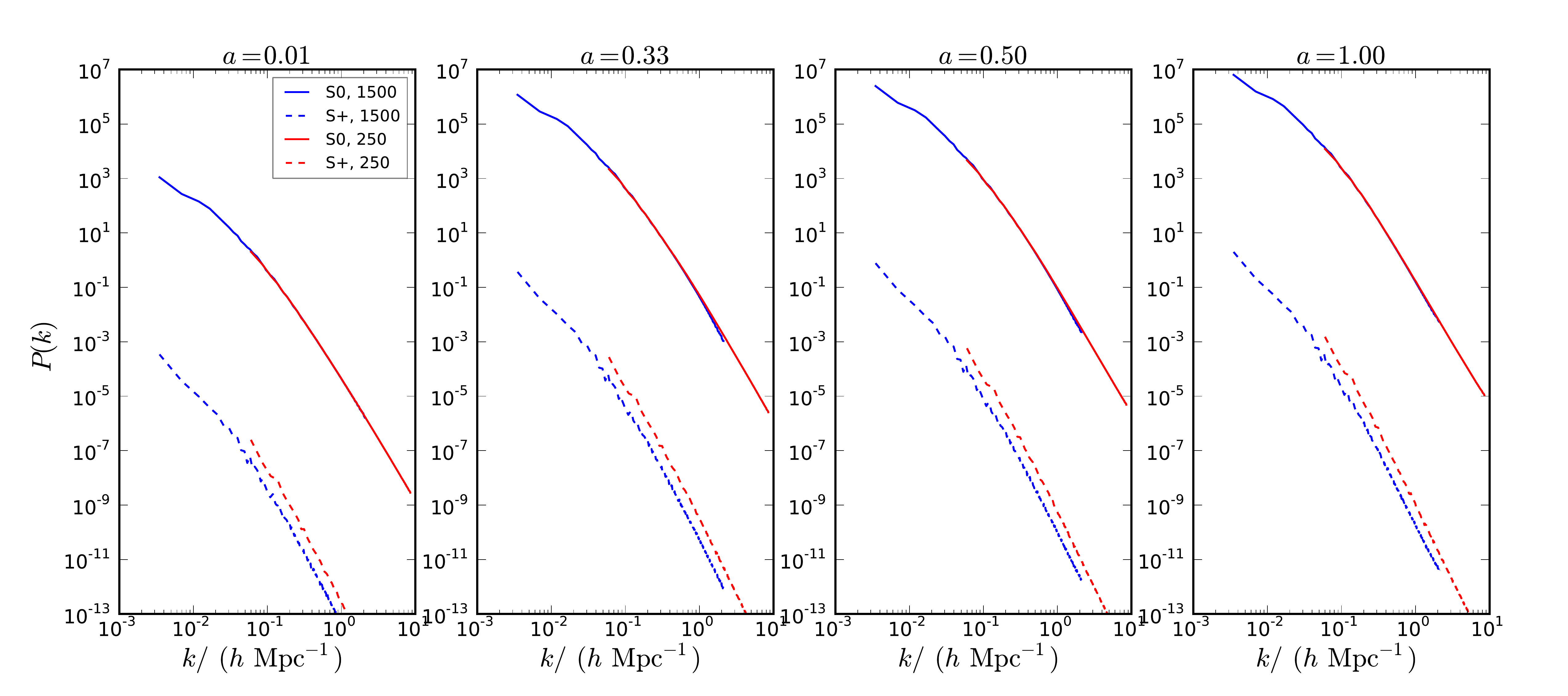}
\caption{ The numerical scalar components of $\bP$ are further broken down into helicity components, the helicity-0 component  (solid line, S0) and helicity-$+$ component (dashed line, S+). Results from two boxes are shown: 1500 $ \MpcOh $ (blue) and  250 $ \MpcOh $ (red). For the scalar components, there should be no helicity-$+$ parts, and the amount of  helicity-$+$ components present suggests the accuracy of the numerical algorithm. The error, \textit{i.e.},~the helicity-$+$ component is six orders of magnitude smaller than the signal.  Note that the helicity-$-$ components are not shown as they are identical to the helicity-$+$ ones.      }
\label{fig:PkPsi_S_SErr}
\end{figure*}

As we mentioned earlier, we can check the accuracy of the decomposition by further breaking down the field in the  helicity basis. For the scalar part, there should be no helicity-$\pm$  components, and the amount of the residual suggests the accuracy of the algorithm. In Fig.~\ref{fig:PkPsi_S_SErr}, we plot the components of the numerical scalar part in the helicity basis. The results are indeed dominated by the helicity-$0$ part. However, there is a small amount of helicity-$+$, but its value is six orders of magnitude smaller than the signal. We do not show the helicity-$-$ part because it is identical to the helicity-$+$ one as expected from symmetry.  Also note that the helicity-$+$ residuals from the two boxes do not overlap, suggesting this may arise from numerical artifacts.

\begin{figure*}[!htb]
\centering
\includegraphics[ width=\linewidth]{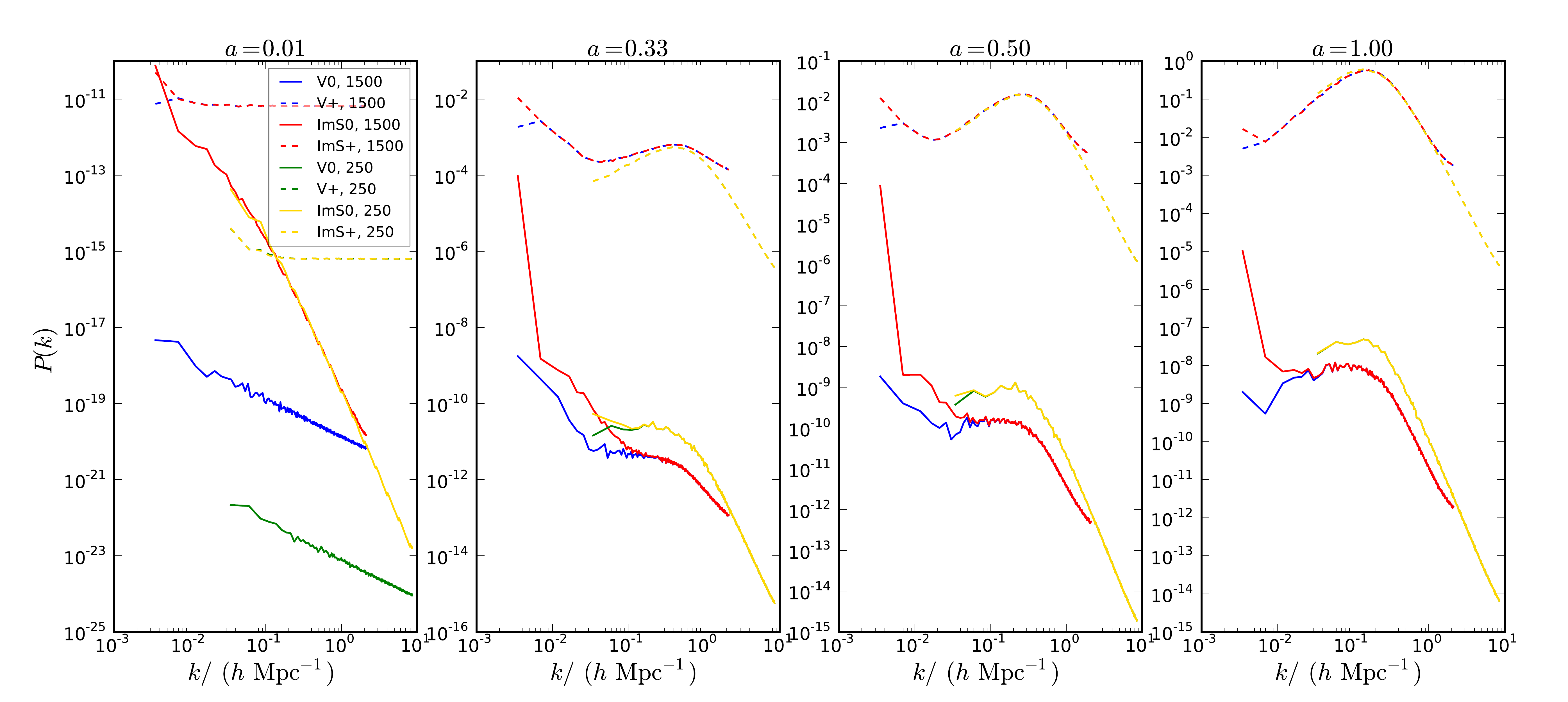}
\caption{ The vector components of $\bP  $ obtained from vector potential (V, blue and green) and subtracting the scalar components from the input field (ImS, red and yellow) are further decomposed into helicity components as a sanity check. The signal is the helicity-$+$ part (dashed), while the amount of helicity-$0$ (solid) component suggests the accuracy of the algorithm.  The smaller box (250 $\MpcOh$) yields more accurate results than the large one (1500 $\MpcOh $) due to much better mass resolution. 
  }
\label{fig:PkPsi_V_VErr}
\end{figure*}

We now go on to look at the vector part of $\bP  $ more carefully. As it is a small quantity, we first try to identify possible spurious numerical artifacts. As for the case of the scalar component of $\bP$, we will further break it down into the helicity components as a sanity check. In Fig.~\ref{fig:PkPsi_V_VErr}, we display the various components of the vector components of $\bP$. At $z=99$, the initial conditions are set by 2LPT, so there should be no vector components of $\bP$ only.  Thus the powers we see are errors. We note that there is some constant residual vector power spectrum.  For the 1500 $\MpcOh $ box, its magnitude is about $10^{-11}   (\MpcOh)^5$, while for the 250 $\MpcOh $   box,  its magnitude is much smaller, about $10^{-15}   (\MpcOh)^5 $. Recall that for the scalar components of $\bP$  in Fig.~\ref{fig:PkPsi_S_SErr}, the results are much less sensitive to the box sizes.  For the vector part the smaller boxes yield much more accurate results than the smaller one.

As redshift decreases, the helicity-$+$ part develops a bump at small scales for $k \sim 0.5 \hOMpc$. For this bump, the results from both boxes agree with each other. However, the results from the large box suggests that the power spectrum goes up as $k$ decreases for $k < 0.02 \hOMpc$. This part of the spectrum is in fact time independent.  Because we expect that the vector power spectrum is generated by shell crossing at small scales, this feature must not be physical. Furthermore, at $z=2$, we note that the helicity-$+$ component from the smaller box differs from the larger box one and  keeps on decreasing as $k$ decreases. The vector components sensitively depend on the  mass resolution, \textit{i.e.} the mass of the particle in the simulation.  The smaller box has much better mass resolution than the larger one.  When the mass resolution is poor, the vector components are spuriously enhanced. Similar results are also found in the context of vorticity \cite{PueblasScoccimarro2009}. Therefore we will not consider the first trough at large scales from the large box any further.  

The two different methods of obtaining the vector components yield very similar results, except for the large box at the  largest scales. Also the error of the decomposition, \text{i.e.}, the helicity-$0$ components are generally small, about six orders of magnitude smaller than the signal. However, the error increases rapidly for the large box for $k \lesssim 0.02 \hOMpc$. All these suggest that the results from the large box at the largest scales are not reliable. 

\begin{figure}[!htb]
\centering
\includegraphics[ width=\linewidth]{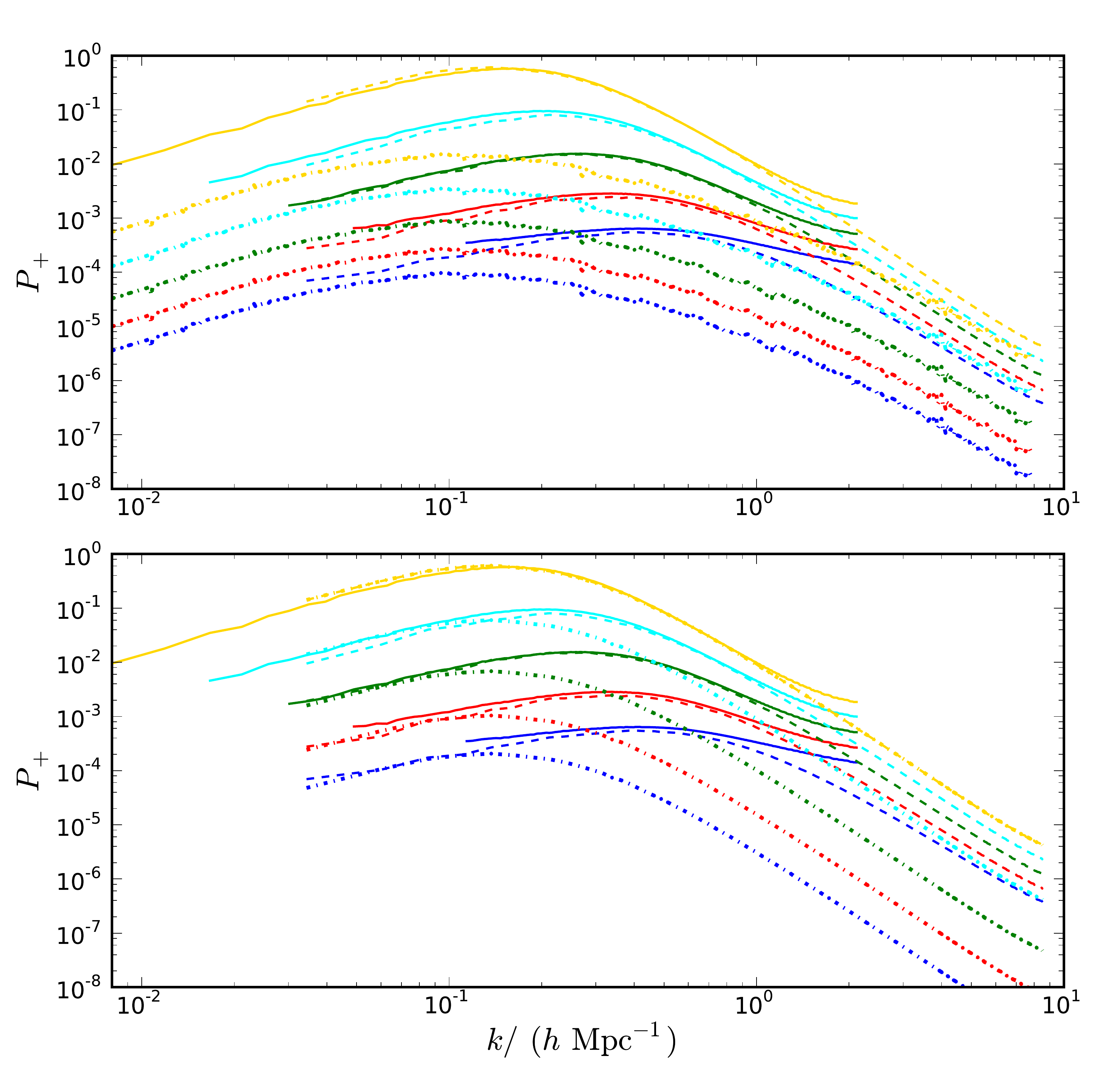}
\caption{    The helicity-$+$ power spectrum of $\bP$. The set of curves are for $z=2$ (blue), 1.5 (red), 1 (green), 0.5 (cyan) and 0 (yellow), respectively (from bottom to top).  The solid curves are from 1500 $\MpcOh$ box and the dashed ones are from  250 $\MpcOh$ one. In the upper panel, we show the power spectrum from the vector contribution in LPT $P^{33 \rm v}  / 2 $ (dotted-dashed).  In the lower panel, the dotted-dashed  curves are obtained by scaling the measurement from the 250 $\MpcOh$ box at $z=0$ by the factor $D^{9.5}$.   Those data points at the large scales from the 1500 $\MpcOh$ simulation with potential spurious artifacts have been removed for clarity.   }
\label{fig:PkV_P33v_scaling}
\end{figure}

We display the helicity-$+$ power spectrum of $\Psi$ for various redshifts in Fig.~\ref{fig:PkV_P33v_scaling}. We show results from both the 1500 and 250 $\MpcOh  $ boxes. However,  as we discussed above, there are potential spurious artifacts in the vector power spectrum at the largest scales in the large box simulation. For the sake of clarity, we have removed the data points at the largest scales in the 1500 $\MpcOh$ box. Those removed data points show an increasing trend as $k$ decreases, and they collapse to the same line at the large scales.  In the intermediate scales, $k \sim 0.06 -0.4 \, \hOMpc $, where both boxes have good resolution, they agree with each other. Around $k\sim 2  \, \hOMpc $, the large box gives higher power than the small box, and this is due to aliasing in the power spectrum of the large box simulation.

The helicity-$+$ contribution from LPT,   $P^{33 \rm v} /2 $, is shown in  the upper panel in   Fig.~\ref{fig:PkV_P33v_scaling} for comparison. At $z=2 $, the signal detected is more than an  order of magnitude greater than the LPT contribution for $k > 0.4 \hOMpc$,  while at large scales, the signal is closer to the LPT results. This is consistent with the picture that a significant amount of vector contribution is generated in shell crossing at small scales.   As redshift decreases,  the power at large scales grows more rapidly than the LPT results.  At $z=0$, the vector contribution from LPT is an order of magnitude smaller than the signal detected for $k\lesssim 1 \hOMpc $.

We measure the scaling of the part of the vector power spectrum before the turn-over as a function of time. As the large box suffers numerical issues at large scales, we only fit to the small box results, up to $k=0.1  \,  \MpcOh$. We find that the large scale vector power spectrum can be fitted by a power law $ D^n$, where the best fit is  $n=9.5$, with 1 $\sigma $ error bar [9.2, 10.0].  In the lower panel of Fig.~\ref{fig:PkV_P33v_scaling}, we also show the vector power spectrum  obtained by scaling the one  from the 250 $\MpcOh  $ box  at $z=0$ using the best fit $n$.   It is clear that the turn-over moves to larger scale as redshift decreases.

In  \cite{PueblasScoccimarro2009}, the vorticity power spectrum of the velocity field is measured (Fig.~3 in \cite{PueblasScoccimarro2009}). Vorticity in the velocity field is also generated by shell crossing. To compare the vector power spectrum of $\bP$  with the vorticity power spectrum in \cite{PueblasScoccimarro2009}, it is useful to clarify the relation and difference between them first. The vorticity $\mb{w}  $ is defined as
\begin{equation}
\mb{w} = \frac{  \nabla_{\mb{x}} \times \mb{u} }{f \mathcal{H} }, 
\end{equation}
where $f=d \ln D / d \ln a $, $\mathcal{H} = d \ln a / d \tau   $ and  $\mb{u}  $ is the comoving velocity $ d \mb{x} / d \tau$. Note that the derivative is with respect to the Eulerian coordinate $\mb{x}$. The quantity that is analogous to $\mb{w}  $ is  $\nabla \times \bP  $. The power spectrum of  $\nabla \times \bP  $ is given by 
\begin{equation}
\sum_{i=j}  \langle  \nabla \times \bP(\mb{k}_1 )_i   \nabla \times \bP (\mb{k}_2 )_j \rangle     = k_1^2 \sum_{i=j}  \langle 
 \bP_i(\mb{k}_1 )   \bP_j (\mb{k}_2 ) \rangle.  
\end{equation}
Thus we should multiply by the vector power spectrum of $\bP$ by $k^2$ when comparing with the vorticity power spectrum. One key difference between $\bP  $ and velocity is that $\bP  $ is always defined at the Lagrangian position $\mb{q}$, while velocity is defined at the Eulerian position $ \mb{x} $. Another key difference is that the velocity field is the derivative of $\bP$ at one instant of time, while $\bP  $ gives the cumulative effects over time.

In \cite{PueblasScoccimarro2009}, it was found that when the mass of particle is large, the vorticity is artificially enhanced. Convergence in the vorticity power spectrum is achieved when the mass of particles is less than  $10^9 \, M_{\odot} h^{-1}$ or so. This is similar to our finding that in the large box the vector power spectrum suffers numerical artifacts at large scales, while the small box with particles mass being  $1.1 \times 10^9 \, M_{\odot} h^{-1}$ seems to  be free of numerical artifacts.  Thus mass resolution plays an important role in the measurement of vorticity.

At $z=0$, we find that the vector power spectrum of $\bP  $ turns over at  $k \sim  0.2  \hOMpc $ in Fig.~\ref{fig:PkV_P33v_scaling}. After multiplying by the factor of $k^2$, the turn-over occurs at   $k \sim  0.3  \hOMpc $, while  the vorticity power spectrum turns over at  $k \sim  1 \hOMpc $.  Another difference from  \cite{PueblasScoccimarro2009} is that the growth of the vorticity power spectrum at the largest scales was found to scale as $D^7 $.

In this section, we have measured the scalar and the vector components of $\bP$.  We find that the scalar power spectrum of $\bP$ is suppressed due to shell crossing. Shell crossing also generated vector part of the power spetrum. However, the generated vector part is still much subdominat compared to the scalar one. Thus the scalar assumption is still valid even after the onset of shell crossing.

\subsection{Modifications of LPT} 
\label{sec:ModificationLPT}
In this subsection, and partly in the next one,  we shall examine two modifications of LPT. In the first approach, we shall incorporate the information that shell crossing causes power suppression in the power spectrum of $\bP$ by modifying the displacement potential. Another approach is to combine LPT with the spherical collapse model \cite{KitauraSreffen2012}.    We shall test how good these phenomenological models perform by checking the density power spectrum at the end of this section.  We will see that these approaches yield limited improvements after the onset of shell crossing.

Since we know the power spectrum of $ \bP $ after shell crossing, we may use this extra information to improve the LPT that used to construct halo catalogs. To do so we will fit the numerical power spectrum by multiplying a suppression factor to the LPT $P(k)$.  The functional form we use is 
\begin{equation}
\label{eq:LPTsPkfit}
P_{\rm LPTs} = \frac{ 1 }{ 1 + \alpha k^n } P_{\rm LPT}, 
\end{equation}
where $P_{\rm LPT} $ is LPT power spectrum, and $\alpha $ and $n$ are free parameters. We call it LPTs and s denotes suppression.  We propose  to modify  $\bP_{\rm LPT}  $ to 
\begin{equation}
\label{eq:LPTsPsi}
\bP_{\rm LPTs} (\mb{k} )  =  \frac{ 1 }{ \sqrt{ 1 + \alpha k^n } } \bP_{\rm LPT}.  
\end{equation}
In practice, we generate the LPTs catalogs by multiplying the factor $ 1 /  \sqrt{ 1 + \alpha k^n }    $ to the LPT potential.  This suppression factor serves to suppress the power in small scales. To some extent, the idea is similar to the truncated ZA \cite{MelottPellmanetal1994}, in which the ZA displacement field is computed using  the power spectrum with power beyond the nonlinear scale removed. 

The functional form  Eq.~\ref{eq:LPTsPkfit} does not fit the numerical power spectrum of $\bP$ in the whole range well. Our goal is to fit the large scale part as well as possible, for example up to $k\sim 0.5 \hOMpc$,  and we often find that the resultant fitting power spectrum underestimates the power in the high $k$ regime. We have tried a few simple functional forms, they show qualitative similar behaviors as Eq.~\ref{eq:LPTsPkfit}.  Worse still, it turns out that even Eq.~\ref{eq:LPTsPkfit} provides a good fit to the numerical power spectrum, for example within 5\% up to $k=0.5 \, \hOMpc $. When the fitting formula is fed into Eq.~\ref{eq:LPTsPsi} to generate the catalog numerically, we find that the power spectrum of $\bP $ from the resulting catalog gives much bigger deviation than the fitting formula Eq,~\ref{eq:LPTsPkfit}. This is not surprising given that shell crossing is a highly nonlinear process.

 We carry out the fitting using the 2LPT and 3LPT power spectrum. 
We fit to the numerical results from the 250 $\MpcOh $  box.  For 2LPT, we find that  for the redshifts available the data can be fitted by 
\begin{equation}
\label{eq:2LPTs}
n=1.8,   \quad    \ln  \alpha(z) = 1.3 + 4.6 \ln D(z),
\end{equation}
where $D(z)$ is the linear growth factor, and it is normalized such that it reduces to the scale factor in the matter-dominated era. The fitting power spectrum agrees with the numerical one within 10\% up to $k=
1 \, \hOMpc  $.  The fit is not very good because the 2LPT power spectrum deviates from the numerical one at quite large scale. For the  3LPT power spectrum we find that across different redshifts, the numerical power spectrum can be fitted by 
\begin{equation}
\label{eq:3LPTs}
n=1.5, \quad  \ln \alpha(z) = 2.1 + 3.5 \ln D(z).   
\end{equation}

 

\begin{figure*}[!htb]
\centering
\includegraphics[ width=\linewidth]{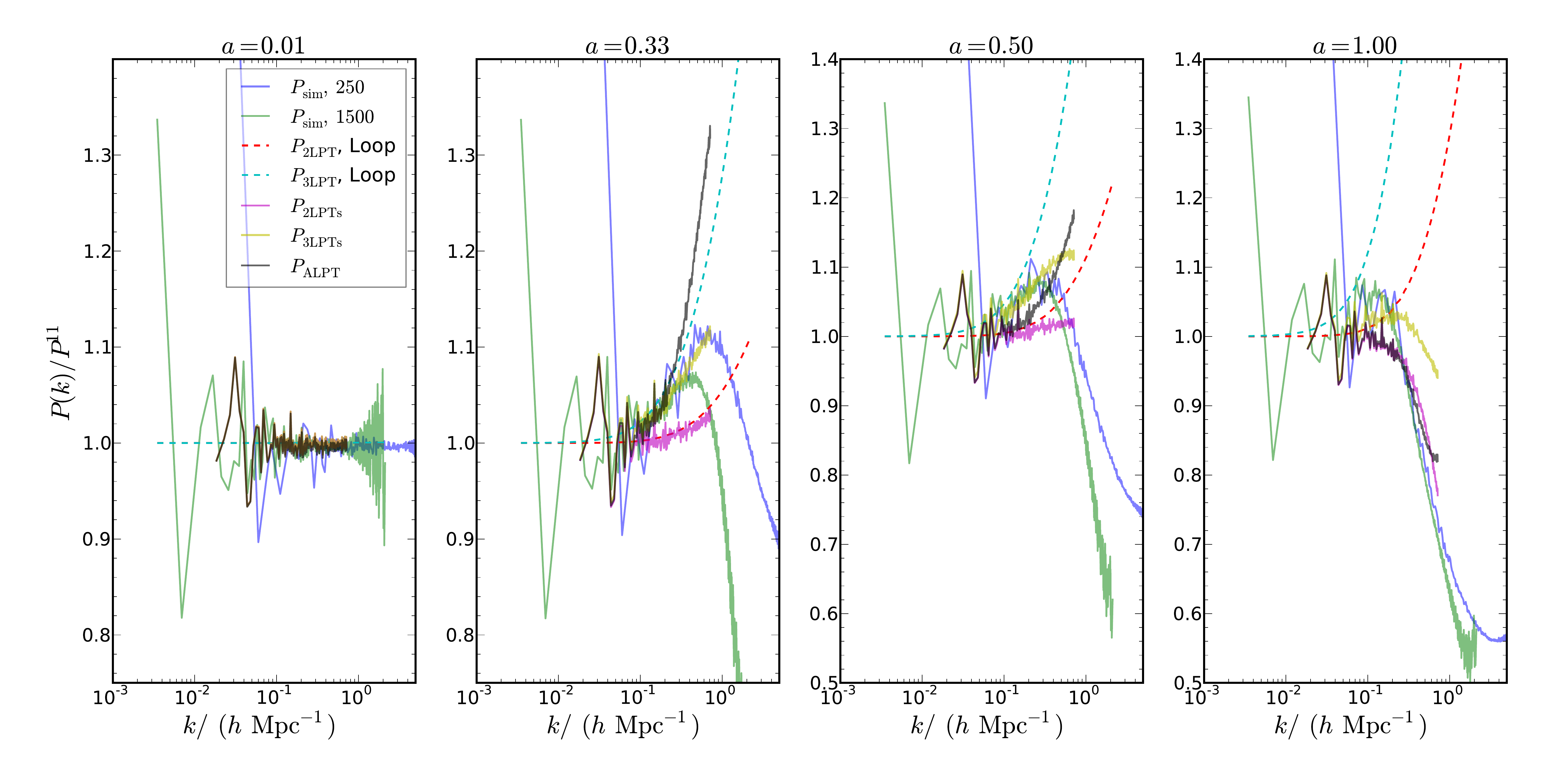}
\caption{ The power spectrum of the $\bP$ obtained from simulation of 1500 (green, solid) and 250 $\MpcOh  $ (blue, solid), 2LPT (red, dashed), 3LPT (cyan, dashed) and 2LPTs catalog  (violet, solid) and 3LPTs catalog (yellow, solid).   We also plot the power spectrum of $\bP$ from ALPT (black, solid) obtained from the prescription Eq.~\ref{eq:ALPT}.    All are normalized with respect to the ZA power spectrum.   }
\label{fig:Pkratio_LPTs_ALPT}
\end{figure*}

We use the best fit Eq.~\ref{eq:2LPTs} and \ref{eq:3LPTs} to generate the LPTs catalogs. The power spectrum of the displacement field from the LPTs catalog is shown in Fig.~\ref{fig:Pkratio_LPTs_ALPT}.   The suppression factor indeed brings the LPT power spectrum much closer to the simulation results. However, we note that the best fit Eq.~\ref{eq:2LPTs} and \ref{eq:3LPTs} fits the simulation results better than those shown in Fig.~\ref{fig:Pkratio_LPTs_ALPT}.  The deviation gets bigger as the redshift decreases.

The fitting formulas Eqs.~\ref{eq:2LPTs} and \ref{eq:3LPTs} are obtained for the standard cosmology parameters. The dependence  on the cosmological parameters have not been checked, although the dependence may be expected to be weak as it is parametrized in terms of the linear growth factor. Also, it may be useful to stress that these formulas are obtained for $\Lambda  $CDM model only, it should be established again for other cosmological models.  However, for our purpose here, we shall use these effective potentials to generate the mock catalogs and see how much we can improve relative to the  standard LPT. We shall use the density power spectrum as the diagnostic.



The second model that we shall examine in detail is a hybrid of LPT and the spherical collaspe model \cite{KitauraSreffen2012}.   Recently, there are some suggestions to improve LPT by reducing the shell crossing using spherical collapse approximation.  References ~\cite{Berbardeau1994, Mohayaeeetal2006} derived a simple evolution equation of Lagrangian volume based on spherical collapse approximation and \cite{Neyrinck2012} found that it agrees well with simulations. But the spherical approximation underestimates the power at large scales. Ref.~\cite{KitauraSreffen2012} then proposed to combine the LPT displacement with the spherical collapse displacement by splitting the displacement vector into large scale and small scale ones. The two regimes are  separated by a filtering scale. When the displacement is smaller than the filtering scale, the displacement field is generated by LPT, while it is given by the spherical collapse displacement for the part that exceeds the filtering scales. The authors called it Augmented LPT (ALPT).  Mathematically, it reads \cite{KitauraSreffen2012}
\begin{equation}
\label{eq:ALPT}
\bP( \mb{k} ) =  W(k,r_s) \bP_{\rm LPT}( \mb{k} ) + [ 1 - W(k,r_s ) ] \bP_{\rm SC }( \mb{k}). 
\end{equation}
In  \cite{KitauraSreffen2012}, $W$ is chosen to be a Gaussian window 
\begin{equation}
W(k,r_s) = e^{-  ( k r_s)^2 / 2  }, 
\end{equation}
and  $r_s=3 \, \MpcOh $ is found to give the best result. The Lagrangian displacement field $  \bP_{\rm LPT}  $ is given by 2LPT and  $ \bP_{\rm SC } $ is obtained from 
\begin{equation}
\nabla \cdot \bP_{\rm sc} =  3  \Big[ ( 1 - \frac{2}{3} D \nabla^2 \phi^{(1)} )^{1/2} - 1 \Big], 
\end{equation}
and if  $1 - \frac{2}{3} D \nabla^2 \phi^{(1)} $ is less than zero, the square root is set to zero.

\begin{figure*}[!htb]
\centering
\includegraphics[ width=\linewidth]{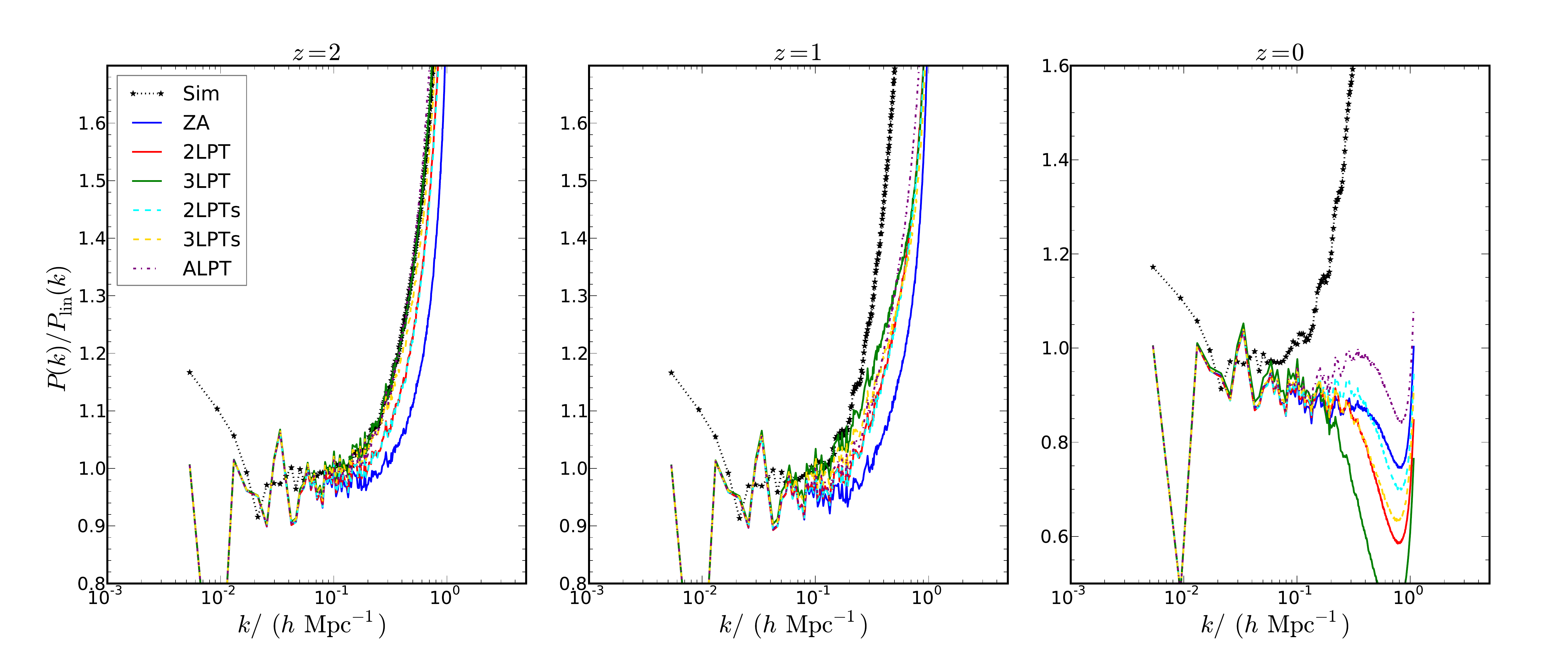}
\caption{ The density power spectrum from simulations (dotted-starred, black) and various LPT recipes: ZA (solid, blue), 2LPT (solid, red), 3LPT (solid, green), 2LPTs (dashed, cyan), 3LPTs (dashed, yellow) and ALPT (dotted-dashed, violet). The subplots are for $z=2$, 1, and 0 respectively (from left to right).  All are normalized with respect to the linear density power spectrum.  }
\label{fig:Pkdelta}
\end{figure*}

   We now proceed to compare the density power spectrum obtained using various recipes with that from simulation. The LPT catalogs are produced and the particles are interpolated to the grid by the Cloud-in-Cell algorithm to compute the power spectrum.  In Fig.~\ref{fig:Pkdelta}, we compare the density power spectrum obtained from various approaches against the simulation results. Note that the random seed used in the simulation is different from the one in the  LPT catalogs, and that is why their power is quite different at large scales.

At high redshift, $z=2$, higher order LPT performs better than the lower order one. 3LPT tracks the $N$-body results well and the power of 3LPT is within 1\% from the $N$-body one within  $k = 0.4 \hOMpc $ shown.  LPTs does not give any better results than the standard LPT. 3LPTs in fact yields slightly lower power than 3LPT, 2LPTs gives almost the same results as 2LPT.  The performance of ALPT is similar to 3LPT, although it  gives slightly higher power than $N$-body results for $ k > 0.5 \hOMpc  $.

As redshift decreases, the differences between LPT results and simulation widen. At the intermediate redshift,  $z=1$, higher order LPT still outperforms the lower order one. 3LPT is still the best in the mildly nonlinear regime,  the power is only 4\% lower than the $N$-body results up to  $k \sim 0.25 \hOMpc $. As at $z=2$, 3LPTs  yields slightly lower power than 3LPT, and 2LPTs performs very similar to 2LPT. We also note that all the LPT recipes cluster around a small strip for $ k > 0.6  \hOMpc $. ALPT yields slightly lower power than 3LPT in the intermediate regime as it is based on 2LPT, however, it gives higher power than 3LPT for $ k > 0.5 \hOMpc$.

 At $z=0$  almost all the LPT recipe results fall below the linear theory one. At the mildly nonlinear regime, for standard LPT, the higher the order of perturbation, the lower the power is.  ZA gives higher power than 2LPT and 3LPT for $ k > 0.3 \hOMpc $.  In the weakly nonlinear regime $ k\sim 0.1 \hOMpc $, LPTs yields slightly higher power than LPT. For $k > 0.3 \hOMpc$. ALPT results in the highest power among all the LPT recipes for $ k > 0.1 \hOMpc $.

Finally, we note that the scales that the LPT results deviate substantially from the simulation results in the density power spectrum is quite similar to that in the power spectrum of $ \bP$. For example, 5 \% deviation of the 3LPT results from the numerical one occur roughly around 0.3 (at $z=1$) and 0.1 $\hOMpc $ (at $z=0$) in both cases.

As far as the density power spectrum in the mildly nonlinear regime is concerned, various LPT recipes still fall short of the simulation results.  Two of its variants examined here do not give any significant improvement for the power spectrum in the mildly nonlinear where standard LPT is known to break down due to severe shell crossing. In particular for  the two variants of LPT considered here, information from $N$-body simulation has been used already. In LPTs, the effective potential is derived from the fitting formulas Eq.~\ref{eq:2LPTs} and \ref{eq:3LPTs}, while in ALPT, the primarily motivation is that the scatter plot of $ \nabla \cdot \bP  $ (see also the next subsection) from spherical collapse model agrees with simulations well \cite{Neyrinck2012}.  While the information is taken from some statistics in which some averaging is taken, given shell crossing is a highly nonlinear process, it may not be surprising that this effective approach would fail for some other statistics, such as the density power spectrum. This suggests that detailed modeling of the small scale physics is required in order to improve the standard LPT.

\subsection{ Scatter plot of $ \nabla \cdot \bP $  }
\label{sec:ScatterDivPsi}

As we see previously the vector part of the displacement field is small,  in this section  we will focus on  $ \nabla \cdot \bP $, which captures all the information if  $\bP_{\rm  } $ is potential.  For both  $  \bP_{\rm fin } $ and  $\bP_{\rm ini } $, the divergence is taken with respect to the Lagrangian coordinate $\mb{q}$.  In this section, we shall explore the information that can be obtained from the scatter plot between  $  \bP_{\rm fin } $ and  $\bP_{\rm ini } $.

\begin{figure*}[!htb]
\centering
\includegraphics[ width=\linewidth]{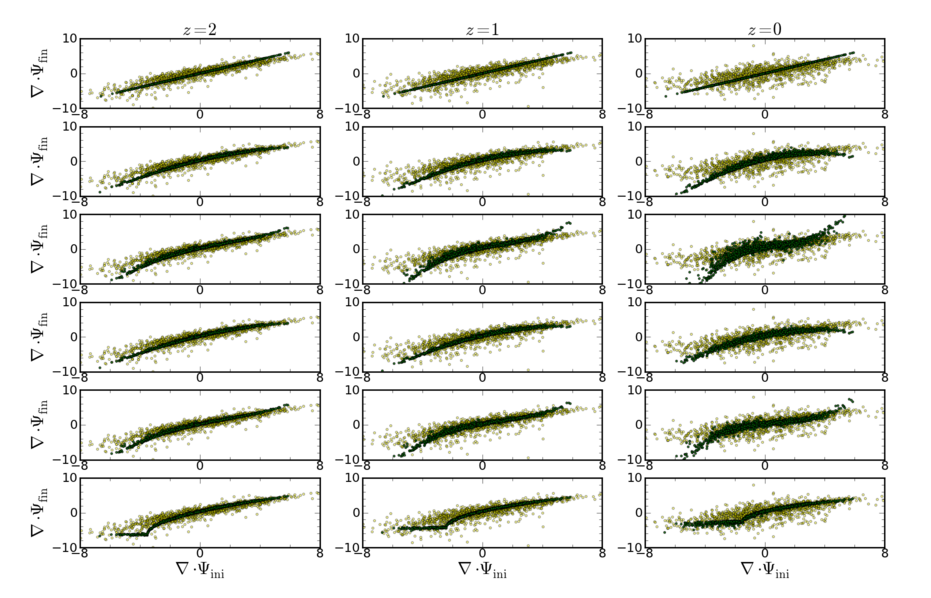}
\caption{The scatter plot between the initial $\nabla \cdot \bP_{\rm ini } $ and the final  $\nabla \cdot \bP_{\rm fin } $. The simulation results are shown as yellow circles. The three columns correspond to $z=2$, 1, and 0, respectively (from left to right). On top of the simulation results, we show the corresponding scatter obtained from ZA, 2LPT, 3LPT, 2LPTs, 3LPTs, and ALPT (green dots, from top to bottom). Both  $\nabla \cdot \bP_{\rm ini } $   and  $\nabla \cdot \bP_{\rm fin } $ have been multiplied by the appropriate linear growth factor $D$ to bring them to $z$=0. }
\label{fig:DivPsi_scatter_theory}
\end{figure*}

We shall first examine various LPT recipes using the scatter plot  of   $ \nabla \cdot \bP $.   In Fig.~\ref{fig:DivPsi_scatter_theory}, we plot  $\nabla \cdot \bP_{\rm ini } $ at the initial time against  $\nabla \cdot \bP_{\rm fin } $ at the final time, as in \cite{Neyrinck2012}.   We have multiplied $\nabla \cdot \bP $ by the linear growth factor $D$ to bring them to $z=0$.  

We compare the scatter plot for $\nabla \cdot \bP $ obtained from simulation against those from the LPT recipes. For  $\nabla \cdot \bP $ from simulation, as redshift decreases, the scatter increases. The relation between  $\nabla \cdot \bP_{\rm ini} $  and  $\nabla \cdot \bP_{\rm fin} $  is linear in ZA. For higher order LPT, such as 2LPT and 3LPT, they depend on other derivatives of the deformation tensor as well, not just its trace, and so there are scatters in the relation.  There is  less scatter in all the LPT recipes than in the simulation.  The mean relation is roughly quadratic for 2LPT and cubic for 3LPT \cite{Neyrinck2012}.   At low redshifts, these behaviors in the positive and negative ends of  $\nabla \cdot \bP_{\rm ini} $ deviate from the simulation markedly. In LPTs, thanks to the suppression factor, the deviations from the simulation results in the ends are reduced, and so the agreement with simulations are improved. We also show the scatter obtained from ALPT. The scatter in ALPT follows the mean of the simulations closely. We also note that the scatter in ALPT is much reduced as in the spherical approximation only the trace of the deformation tensor appears. The spherical collapse model tracks the mean of the scatter plot well was in the original motivation for ALPT \cite{Neyrinck2012,KitauraSreffen2012}.

Ref.~\cite{Neyrinck2012} reported that there are some differences between the scatter plot constructed using the FFT and FD methods.  In \cite{Neyrinck2012}, the derivatives were done using the FD method, and found that there is an accumulation of points around $\nabla \cdot \Psi_{\rm fin} = -3$, where $\Psi_{\rm fin}  $ is a physical displacement field without the linear extrapolation factor.  \cite{Neyrinck2012} also pointed out, when spectral derivatives are used, there is no saturation around $ -3 $.  Given the better precision in reconstruction for the FFT method described in Appendix \ref{sec:TestCases}, we use the spectral derivatives here.

In the rest of the section, we shall explore the information about the various kinds of objects in the scatter plot.  In LPT, the Eulerian density is obtained from the mass conservation equation as 
\begin{equation}
1 + \delta(\mb{x} ) = \frac{ 1  }{J } ,   
\end{equation}
where $J$ is the Jacobian determinant 
\begin{equation}
J =  \det \Big(  \frac{ \partial \mb{x} }{  \partial \mb{q}    }  \Big) .
\end{equation}
In ZA, $J$ is given by 
\begin{equation}
J = ( 1 - D \lambda_1 ) ( 1 - D \lambda_2 ) ( 1 - D \lambda_3 ),  
\end{equation}
where  $\lambda_i $ is the eigenvalue of the deformation tensor $\nabla_{ij} \phi^{(1)} $, and they are ordered such that $\lambda_1  \geq \lambda_2 \geq \lambda_3 $.  By examining the eigenvalues of the deformation tensor, one can classify the collapsed structures. The vanishing of the factor in $J$ implies that the axis associated with that eigenvalue has collapsed. We assume that all the cosmic structures can be classified into knots (3 collapsed axes), filaments (2 collapsed axes), sheets (1 collapsed axis) and voids (no collapsed axis).  We can set some cuts on the eigenvalues of $\nabla \cdot \bP_{\rm ini } $ to select some objects, and explore how they distribute in the scatter plot.

It is important to point out that this kind of classification is based on the analysis at one scale only.  As pointed out in \cite{LeeShandarin1998} for the case of collapsed objects, this analysis suffers from the cloud-in-cloud problem in the Press-Schechter argument. That is the identified local structure may be hosted within the structure of another kind. Thus objects identified here may not agree with those from the more sophisticated identification algorithms (see for example \cite{Cautunetal2013,TempelStoicaetal2013} and references therein). Indeed, to get the right abundance of collapsed objects,  a large factor is required \cite{LeeShandarin1998}. With this caveat in mind, we shall use this simple classification here.

 In Table \ref{tab:CollaspedStructure}, we show the classification of the collapsed structure using the eigenvalues of $J$.  This is based on the analysis of the field at the scale of the grid  $\sim 1.5 \MpcOh$.  For example, the condition that $\lambda_1 \geq t =  1/D  $ means that at least 1 axis has not collapsed. These kinds of objects include filaments, sheets and voids. The probability distribution of  the ordered $\lambda_i$ for a Gaussian field has been calculated \cite{Doroshkevich1970,LeeShandarin1998}. By integrating the probability distribution (Eq.~13, 14, and 15 in \cite{LeeShandarin1998}) from the threshold $t$ to infinity we get the fractions in Table \ref{tab:CollaspedStructure}.  In Eq.~13, 14, and 15 in \cite{LeeShandarin1998}, there is a free parameter $\sigma$, the rms variance of the density. Using the  $\sigma$ obtained by computing the variance of  $\nabla^2 \phi^{(1)} $ at the grid, we find that the fraction computed agrees with the direct measurements very well, within 0.5\%. We note that more than 99\% of the Lagrangian volume belongs to the group containing sheets, filaments and voids. That is the Lagrangian volume that collapse to form halos are less than 1 \%. As redshift decreases,  the fraction of cosmic voids decreases, while that of the sheets and filaments increases. From Table \ref{tab:CollaspedStructure}, we deduce that  at $z=0$, 0.8\% of the Lagrangian volume forms knots, 15\% forms sheets, 51\% forms filaments and 33\% forms voids. This is in line with visual expectation in the large scale structure that the cosmic web is dominated by filaments and voids.

\begin{table*}
\caption{ Fractions of the Lagrangian volume that form various large scale structures. Classification of collapsed objects based on the eigenvalues, $\lambda_i  $, of the deformation tensor. $t$ is the threshold $1/ D$.  }
\label{tab:CollaspedStructure}
\begin{ruledtabular}
\begin{tabular}{ |l|l|l|l| }
                     &  sheets, filaments and voids   &   filaments and voids                 &    voids                                                          \\ 
                     &  $ \lambda_1 \geq t      $   &  $ \lambda_1 \geq  \lambda_2 \geq t$  &    $ \lambda_1 \geq  \lambda_2 \geq  \lambda_3   \geq t$          \\
\hline 
 $z=2$  &  0.9999                     &   0.990                               &        0.79           \\
 $z=1$  &  0.9986                     &   0.946                               &        0.56           \\
 $z=0$  &  0.9917                     &   0.842                               &        0.33           \\
\end{tabular}
\end{ruledtabular}
\end{table*}

\begin{figure}[!htb]
\centering
\includegraphics[ width=\linewidth]{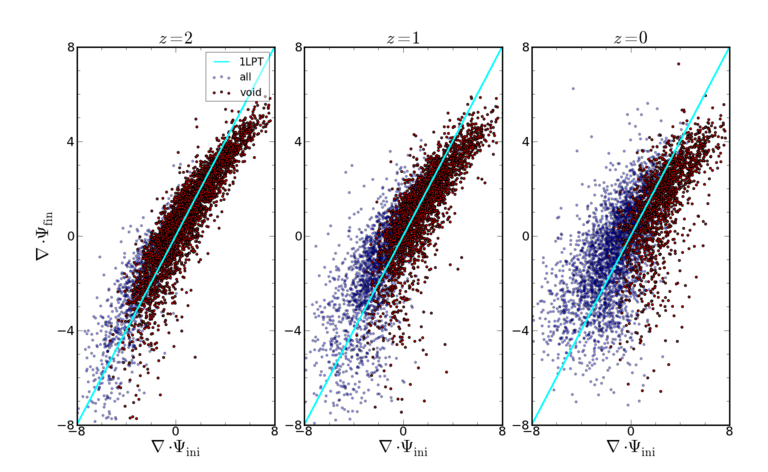}
\caption{  The scatter plot of voids (red) on top of the simulation results (blue). The cyan line corresponds to the 1LPT result.  
     }
\label{fig:DivPsi_scatter_V}
\end{figure}

 In Fig.~\ref{fig:DivPsi_scatter_V}, we show the scatter plot of $\bP $ for voids on top of the full simulation results. The voids mostly distributed around the positive side $\nabla \cdot \bP_{\rm ini } $. As the redshift decreases,  the fraction decreases and they move towards the more positive end of  $\nabla \cdot \bP_{\rm ini } $.  Since voids are the region that has not yet collapsed, one may think that they undergo less shell crossing than some arbitrary region. This idea is borne out in Fig.~\ref{fig:DivPsi_scatter_V}. The scatter in the full simulation decreases as  $\nabla \cdot \bP_{\rm ini } $ increases, and the void region corresponds to the positive end of  $\nabla \cdot \bP_{\rm ini } $.

\begin{figure}[!htb]
\centering
\includegraphics[ width=\linewidth]{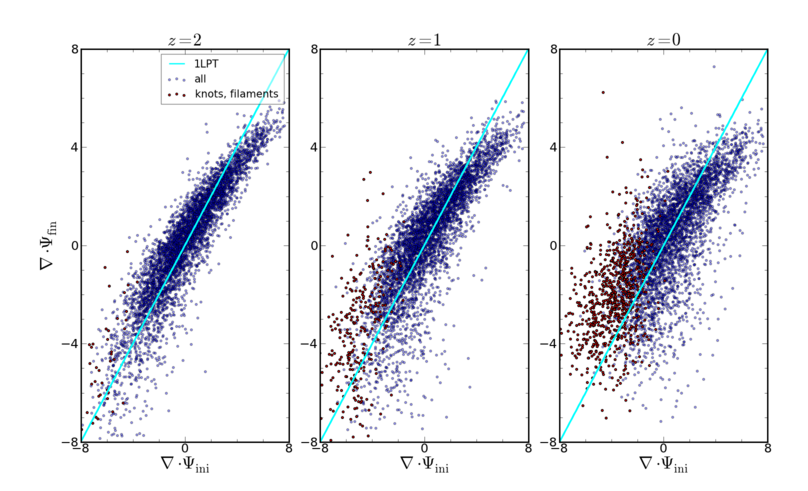}
\caption{  The scatter plot of knots and filaments (red) on top of the simulation results (blue). The cyan line corresponds to the 1LPT result.
     }
\label{fig:DivPsi_scatter_HF}
\end{figure}

In Fig.~\ref{fig:DivPsi_scatter_HF}, we show a similar plot for knots and filaments. These collapsed structures mainly distribute around the negative end of  $\nabla \cdot \bP_{\rm ini } $. As redshift decreases, the region expands from the negative  $\nabla \cdot \bP_{\rm ini } $ end to the positive side. These collapsed objects are virialized and have undergone more shell crossing. They show less dependence on the initial $\nabla \cdot \bP_{\rm ini } $, as manifested with larger scatter.

\section{Conclusions}
\label{sec:Conclusions}
Lagrangian displacement field $\bP$ is the central object in LPT. LPT is very successful at high redshifts, but it performs poorly at low redshifts due to severe shell crossing. After shell crossing, the standard LPT breaks down. 

In order to gain insight into  $\bP$ when shell crossing is not negligible, we measure  $\bP$ from $N$-body simulation directly  in this paper. As $\bP$ is potential in LPT to a very good approximation, and shell crossing can generate non-negligible amount of vorticity, we decompose $\bP$  into scalar and vector parts. We use the power spectrum of $\bP$ to quantify the effect of shell crossing. We find that at large scales, the numerical results agree well with 1-loop LPT calculations. However, shell crossing becomes important at low redshifts, and the agreement deteriorates quickly.  At $z=1$,  the 1-loop power spectrum of $\bP$ is about 10\% higher than the results from numerical $\bP$  at around $k\sim  0.3\, \hOMpc $, and it occurs at $k\sim  0.1 \, \hOMpc $ at $z=0$. This is consistent with the more well-known results that the LPT density power spectrum at low redshifts yields much lower power than the $N$-body results due to serious shell crossing.

 We also detect the generation of the vector mode due to shell crossing, although its magnitude is still much smaller than the scalar mode in the mildly nonlinear regime. Our results show that the potential approximation is still good even when shell crossing is non-negligible in the mildly nonlinear regime.  Note that there is vector contribution in third order LPT. The leading contribution from the vector part of LPT to the power spectrum of $\bP$ is of 2-loop order. We find that this 2-loop contribution is much smaller than the signal we detected.   For example, at $z=1$,  the vector power spectrum of $\bP$ contributes to 10\% of the total power spectrum at $k \sim 2.5 \, \hOMpc $,   and this happens at   $ k \sim 1  \, \hOMpc $ at $z=0$. The LPT contribution at these scales are about an order of magnitude smaller than the signal detected. Also the scaling of the large scale vector power spectrum is found to scale as $ D^{9.5} $, while the LPT vector contribution is expected to scale as $D^6$.

We examined the  standard LPT recipes and two of its variants.  In one of the variants, we  incorporate the information of the power spectrum of $\bP$ to improve the generation of catalogs with LPT. We include a power suppression factor to the displacement potential, and the functional form of the suppression factor is obtained by fitting to the numerical power spectrum of $\bP$.  The suppression factor can reduce the deviation from simulations in the void and overdense regions as can be seen from the scatter plot between $\nabla \cdot \Psi_{\rm ini}   $ and $\nabla \cdot \Psi_{\rm fin} $. We used the density power spectrum, which is one of the most important physical observables,  to gauge the performance of LPT and its variants.   However, various LPT recipes still yield power much lower than simulations at redshift close to 0.  The LPT variants yield limited success compared to the standard ones after the onset of shell crossing, even though some information from $N$-body simulation  has been incorporated in the variant of LPT. This is not very surprising given that shell crossing is a highly  nonlinear process. The information is obtained by taking the average for some statistics, it is not guaranteed that other statistics, such as the density power spectrum, will be right.  Our exercises indeed suggest that this is not the case. To improve the standard LPT,  this points to the need for more detailed modeling beyond the simple phenomenological approach.

\section*{Acknowledgment} 
I thank Vincent Desjacques, Cornelius Rampf, Roman Scoccimarro and Xin Wang and the anonymous referee for commenting on the draft of the paper.  I also thank Vincent Desjacques for providing the simulation data used in this work.  This work is supported by the Swiss National Science Foundation.

\appendix

\section{ General structure of the power spectra of $\bP$ } 
\label{sec:GeneralPsiPk}
In this section, we show the general structure of the power spectrum of $\bP $ expressed in terms of its scalar and vector components.

In Fourier space, the Helmholtz decomposition of $\bP $ reads
\begin{equation}
\label{eq:SpectralDPsi}
\bP(\mb{k} )=  \bP_{\rm S}(\mb{k} )  +  \bP_{\rm V}(\mb{k} ),
\end{equation}
with the scalar and vector parts given by 
\begin{eqnarray}
\bP_{\rm S}(\mb{k} )  &= &  i \mb{k} \Phi(\mb{k}) ,   \\
\bP_{\rm V}(\mb{k} )  &= &   i \mb{k} \times \mb{A}(\mb{k} ). 
\end{eqnarray}

The power spectrum between the scalar parts is 
\begin{equation}
\langle \bP_{\mathrm{ S} i }(\mb{k}_{1} ) \bP_{\mathrm{S} j}(\mb{k}_{2} )      \rangle = 
 \mb{k}_{1i}  \mb{k}_{1j} P_{ \Phi} ( k_1 ) \Ddel( \mb{k}_{12}) , 
\end{equation}
where $ P_{ \Phi} $ is the power spectrum of $\Phi$. 

The power spectrum between the scalar part and the vector parts is given by 
\begin{equation} 
\label{eq:crossPhiA}
\langle  \bP_{\mathrm{ S} h }(\mb{k}_{1} ) \bP_{\mathrm{V} i }(\mb{k}_{2} )   \rangle = -  \mb{k}_{1h} \epsilon_{ijk} \mb{k}_{2j} \langle \Phi( \mb{k}_1 ) \mb{A}_k( \mb{k}_2 ) \rangle. 
\end{equation} 
For a statistically isotropic and homogeneous field, the power spectrum can be written in the form 
\begin{equation}
\label{eq:PPhiAcross}
\langle \Phi( \mb{k}_1 ) \mb{A}_l( \mb{k}_2 ) \rangle =  P_{\Phi \mb{A} } ( k_1 ) \mb{k}_{1l}  \Ddel( \mb{k}_{12} ).   
\end{equation}
Using Eq.~\ref{eq:PPhiAcross} we see immediately that the cross power Eq.~\ref{eq:crossPhiA} vanishes. 

The power spectrum between the vector parts is given by 
\begin{equation}
 \langle   \bP_{\mathrm{ V} i }(\mb{k}_{1} ) \bP_{\mathrm{V} l }(\mb{k}_{2} )     \rangle 
 =  - \epsilon_{ijk} \epsilon_{lmn} k_{1j} k_{2m} \langle A_{k} ( \mb{k}_1 )  A_{n} ( \mb{k}_2 ) \rangle   .
\end{equation}
Again for a statistically isotropic and homogeneous field, we have 
\begin{eqnarray} 
  \langle \mb{A}_j (\mb{k}_1 )  \mb{A}_k (\mb{k}_2 )  \rangle & = & \big(  P_{1\mb{AA}} ( k_1 ) \delta_{jk}  \nonumber   \\
& +&   P_{2\mb{AA}} ( k_1 ) \hat{\mb{k}}_{1j} \hat{\mb{k}}_{2k} \big)  \Ddel( \mb{k}_{12})  .
\end{eqnarray}
The condition that $\mb{A}$ is transverse further requires $P_{1\mb{AA} }=P_{2\mb{AA} } $.  
Thus we have
\begin{eqnarray}
 &  & \langle   \bP_{\mathrm{ V} i }(\mb{k}_{1} ) \bP_{\mathrm{V} l }(\mb{k}_{2} )    \rangle  \nonumber    \\ 
& = &    ( k_1^2 \delta_{il} - k_{1i} k_{1l} )   P_{1\mb{AA}} ( k_1 ) \Ddel( \mb{k}_{12} ). 
\end{eqnarray}
Thus the power spectrum of $\bP  $ can be expressed as 
\begin{equation}
\sum_{i=j} \langle  \bP_i( \mb{k}_1 ) \bP_j( \mb{k}_2 ) \rangle = k_1^2 \big(  P_{\Phi} (k_1 ) +  2 P_{1  \mb{A} \mb{A} }(k_1)   \big)  \Ddel( \mb{k}_{12} ).
\end{equation}

\section{ Testing the Helmholtz decomposition algorithm} 
\label{sec:TestCases}

\begin{figure}[!htb]
\centering
\includegraphics[ width=\linewidth]{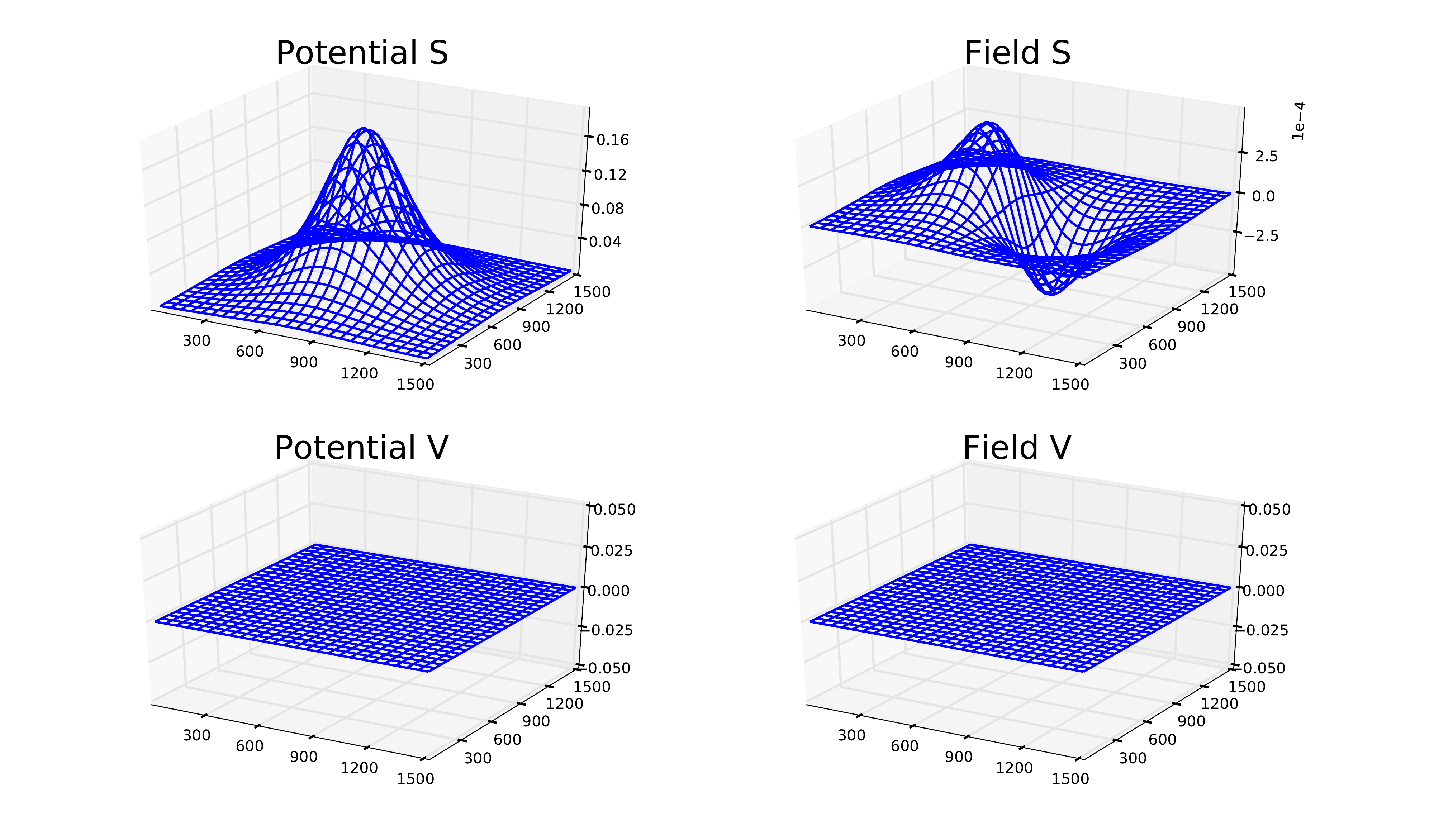}
\caption{  Sections of the scalar part and vector part of the potentials (Eq.~\ref{eq:Test1_Phi} and \ref{eq:Test1_A}) (first column) and its corresponding fields (second column). For vector field, the $x$-component is shown.    }
\label{fig:CongfigPlot_M1_input}
\end{figure}

To test the performance of the decomposition algorithm and the code, we shall consider a couple of examples. The first test is purely longitudinal, and it is given by the potentials
\begin{eqnarray}
\label{eq:Test1_Phi}
\Phi( \mb{r} )        &=& a \mathrm{e}^{ - r'  / \rho} , \\
\label{eq:Test1_A}
\mb{A} ( \mb{r} ) &=& \mb{0}, 
\end{eqnarray}
where  $r' = | \mb{r} - \mb{r}_0 |$,  and   $a$,  $\mb{r}_0 $ and $ \rho $ are some free parameters. From Eq.~\ref{eq:Test1_Phi} and \ref{eq:Test1_A}, we get the vector field 
\begin{equation} 
\label{eq:ScalarVec}
\mb{V} ( \mb{r} )= - \frac{ \mb{r}'  }{ \rho r'  } \Phi.  
\end{equation} 
In Fig.~\ref{fig:CongfigPlot_M1_input}, we show the input potentials and the corresponding fields derived from the potentials. In this plot we have used a box of size 1500 $\MpcOh $ and  a grid of size $120^3 $. The parameters  $a=1$, $ \rho = 150 \MpcOh$ are used, and  $\mb{r}_0 $ is set to the center of the box.  The section is output at constant $z$-coordinate surface with $z=500 \MpcOh $. For vector field, we show the $x$-component of the field.  

\begin{figure}[!htb]
\centering
\includegraphics[ width=\linewidth]{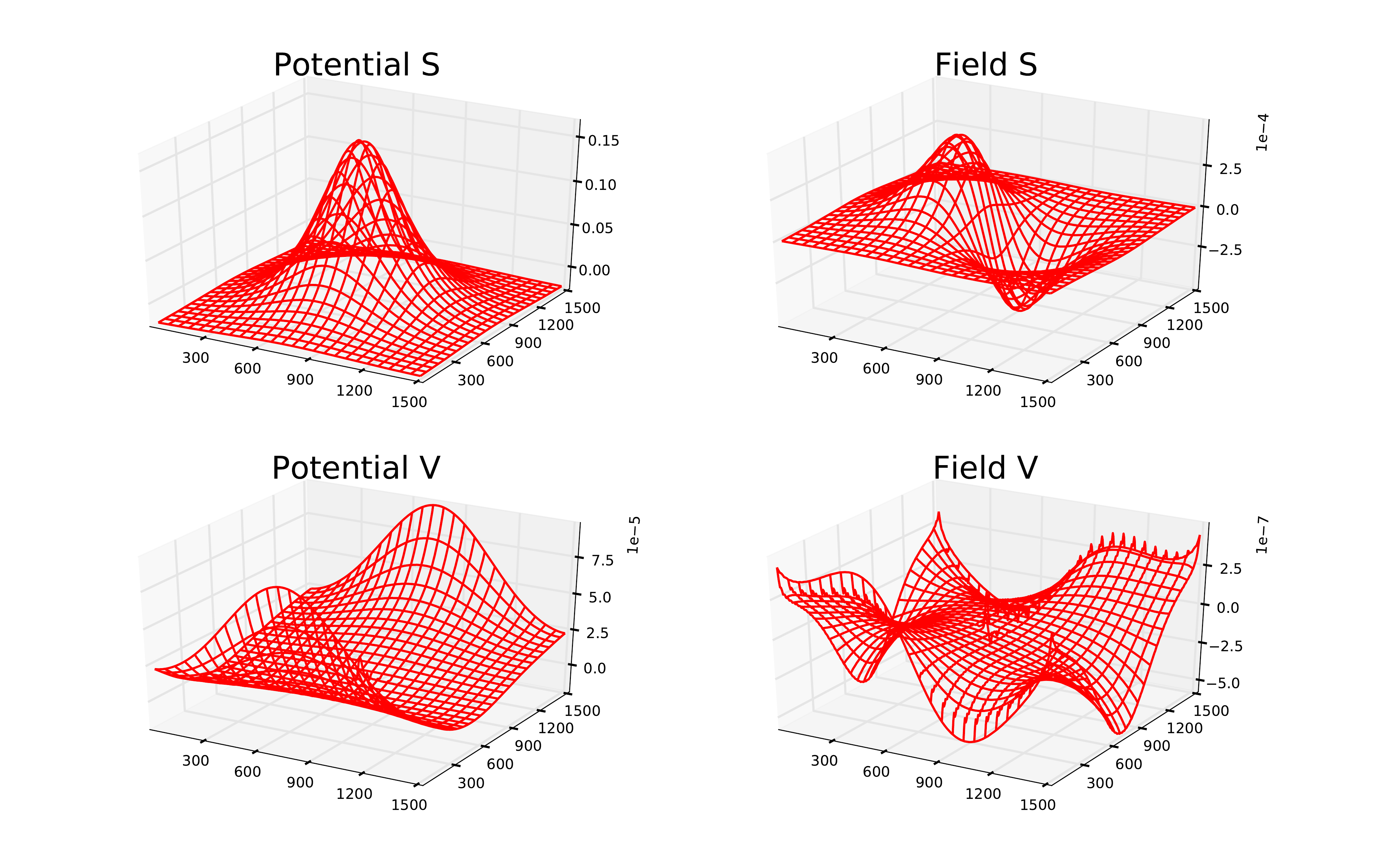}
\caption{  Same as Fig.~\ref{fig:CongfigPlot_M1_input}, but reconstructed using the FFT method.   }
\label{fig:CongfigPlot_M1_reconst_FFT}
\end{figure}

In Fig.~\ref{fig:CongfigPlot_M1_reconst_FFT}, we show the potentials and the field reconstructed using the FFT method. In the FFT method, we compute the divergence and curl of $\bP  $ in Fourier space by the spectral derivatives. We note that the error, that is the vector part of the field, is of three orders of magnitude smaller than the input scalar signal. 

\begin{figure}[!htb]
\centering
\includegraphics[ width=\linewidth]{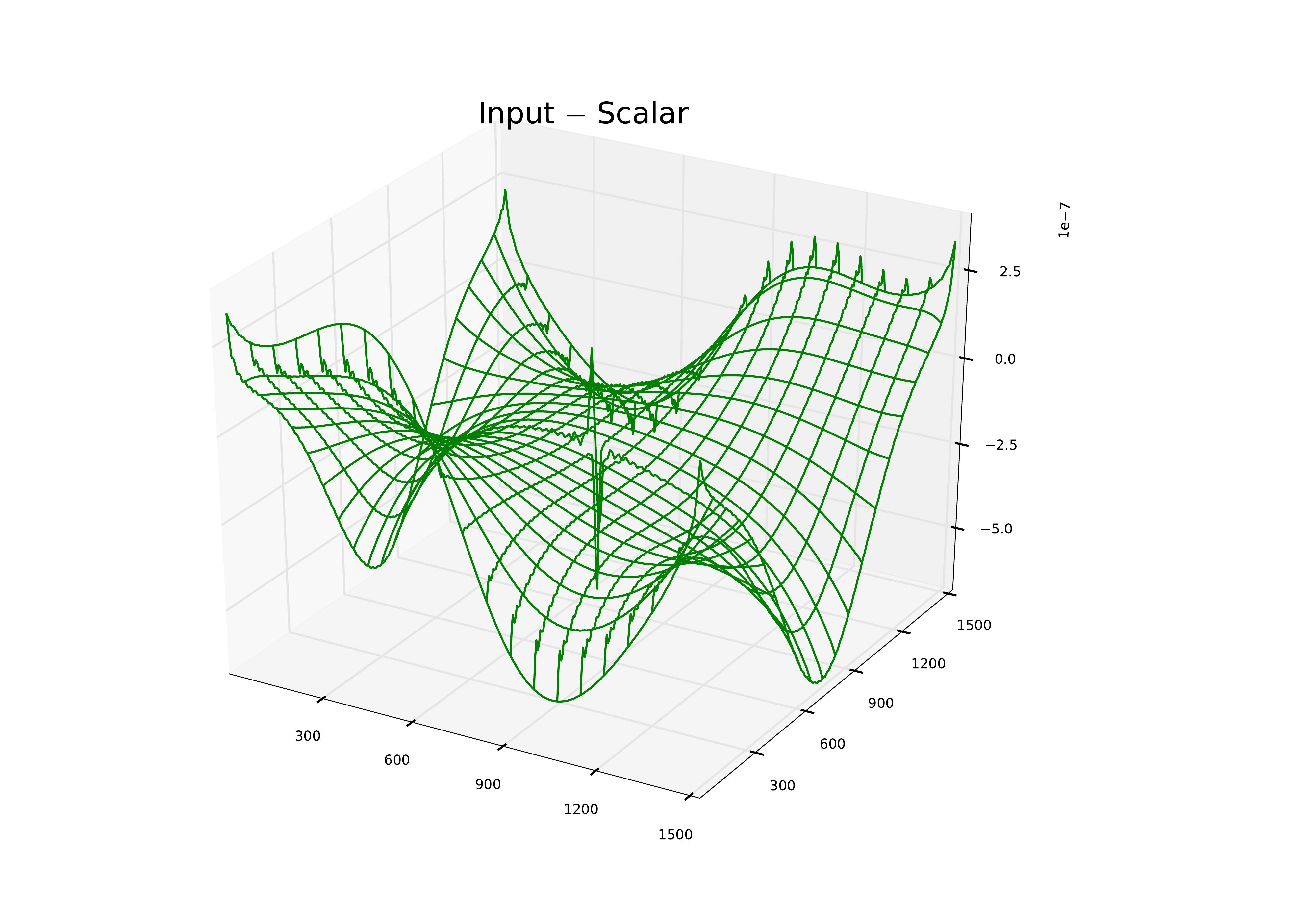}
\caption{ The vector field reconstructed by subtracting the scalar field from the input field. The result is very similar to the one derived from the  vector potential shown in Fig.~\ref{fig:CongfigPlot_M1_reconst_FFT}.     }
\label{fig:Input_Scalar_M1_FFT_single}
\end{figure}

 As we argued previously, in the case of the periodic boundary condition and the field having zero spatial mean, the Helmholtz decomposition is unique. Eq.~\ref{eq:ScalarVec} indeed has zero spatial mean.  Thus, besides reconstructing the vector field by solving the Poisson Eq.~\ref{eq:Poisson_A}, we can obtain the vector part by simply subtracting the scalar part from the input field. We show the reconstruction of the vector field in this way in Fig.~\ref{fig:Input_Scalar_M1_FFT_single}. The result is very similar to that shown in Fig.~\ref{fig:CongfigPlot_M1_reconst_FFT}.

We also do a similar analysis using the finite difference (FD) method. In the FD method, we compute the divergence and curl  of $\bP$ in real space. We apply the symmetric finite differencing method to the nearest neighboring points. We stress that it is important to use a symmetric scheme. In a symmetric finite differencing scheme, the separation between the two points where the difference is made, is two times that of the grid scale. One may think that using an asymmetric scheme, the separation is just one unit of the grid scale, and so a better resolution is achieved. In fact, \cite{Neyrinck2012} used a real space estimator because the author claimed that the real space estimator has twice the resolution of the Fourier space one. We have checked that using this naive asymmetric scheme, the error is larger than the symmetric one by two orders of magnitude. This in fact is consistent with the Nyquist theorem, which states that the highest mode that can be resolved with a grid of size $n_{\rm grid} $ is $k_{\rm F} n_{\rm grid} / 2 $, with $k_{\rm F} $ being the fundamental mode. The highest resolution in real space for a grid of size $n_{\rm grid}$  is the same as that in Fourier space.

\begin{figure}[!htb]
\centering
\includegraphics[ width=\linewidth]{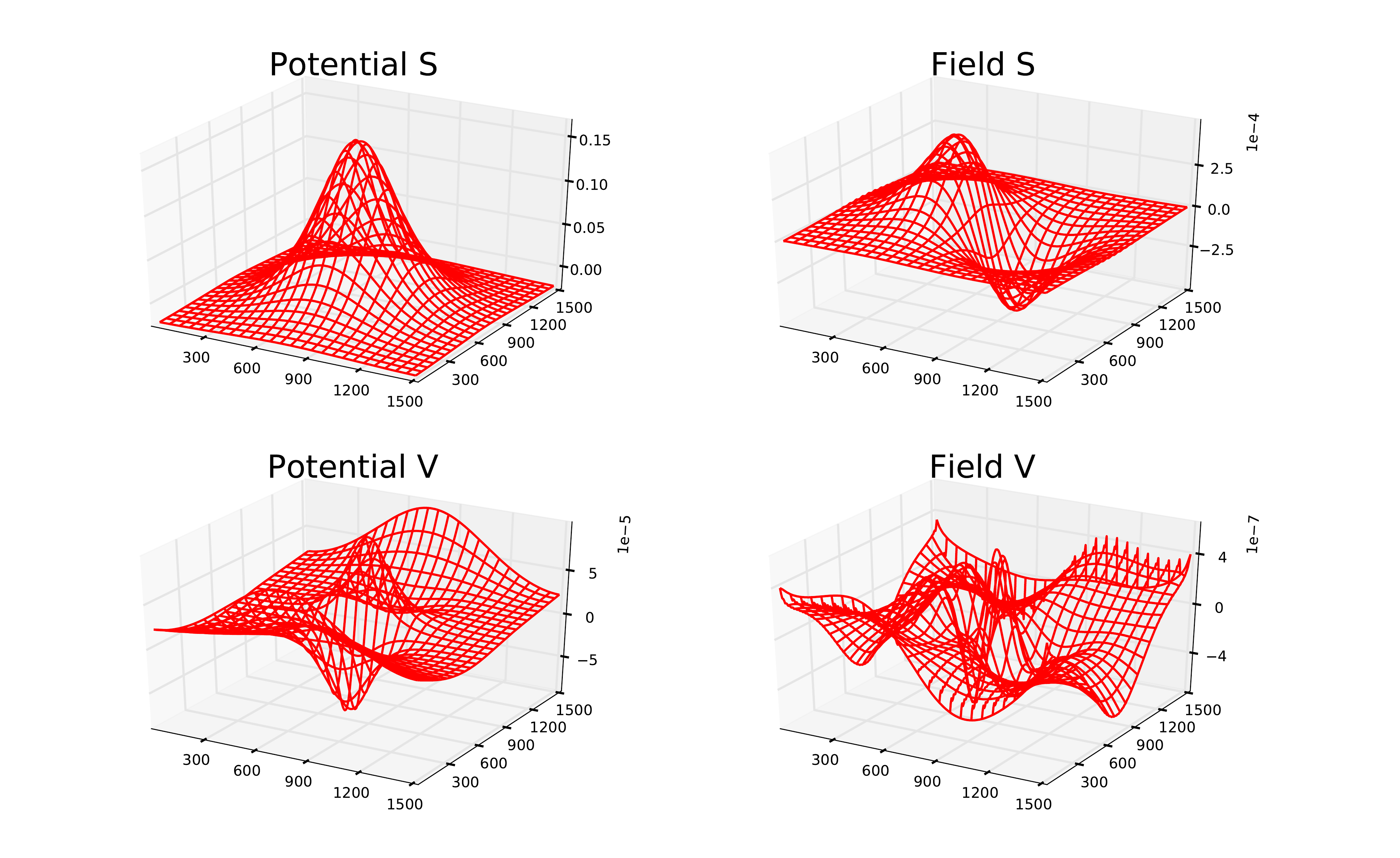}
\caption{  Same as Fig.~\ref{fig:CongfigPlot_M1_input}, but reconstructed using the FD method. The error is larger than that in Fig.~\ref{fig:CongfigPlot_M1_reconst_FFT} by a factor of 2 or so.   }
\label{fig:CongfigPlot_M1_reconst_FD}
\end{figure}

In Fig.~\ref{fig:CongfigPlot_M1_reconst_FD}, we show the results obtained using the FD method. We note that the vector contamination from the FD method is about twice larger than the FFT method. There are also larger fluctuations in the vector contamination in the FD method.

\begin{figure}[!htb]
\centering
\includegraphics[ width=\linewidth]{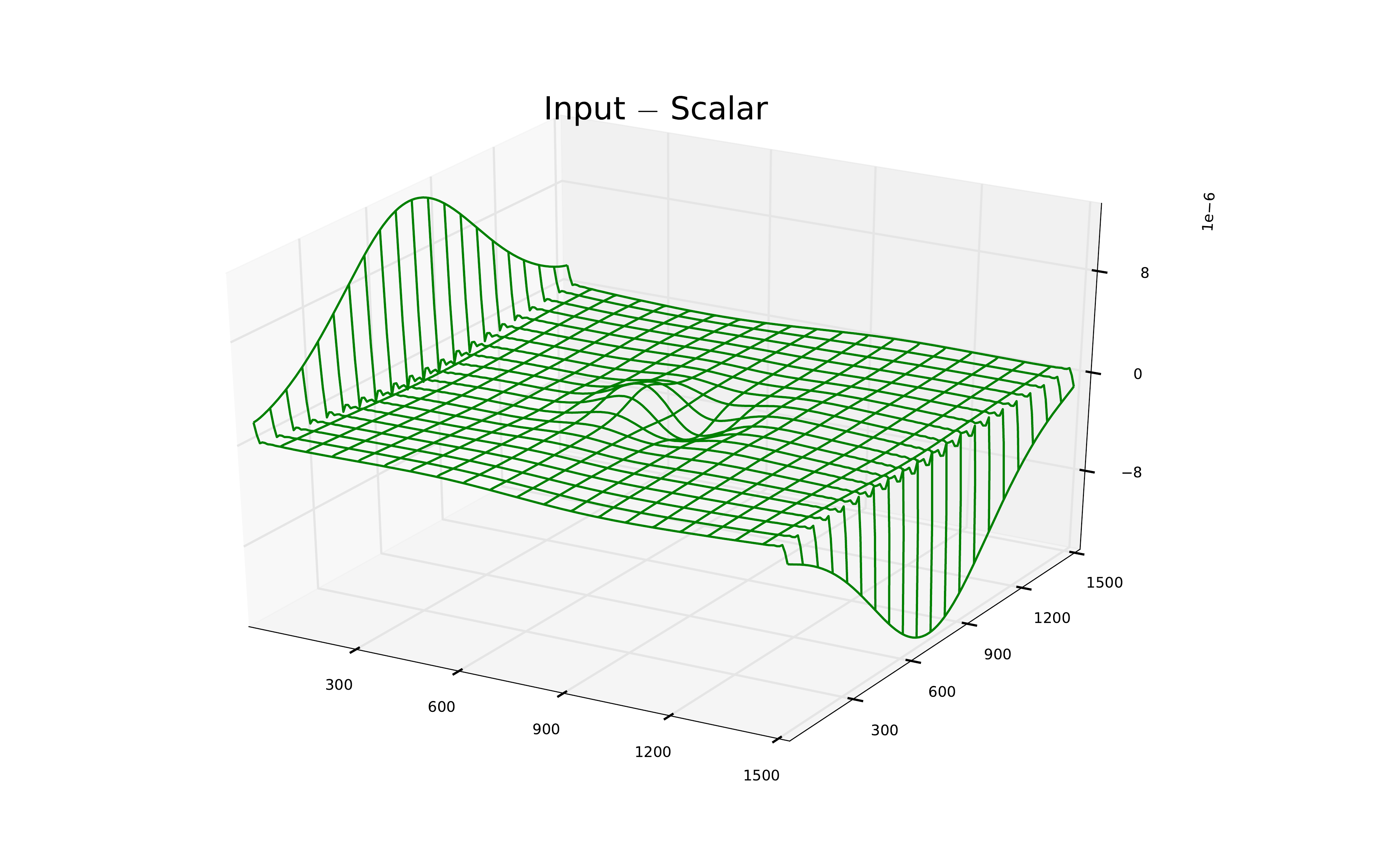}
\caption{ The vector field reconstructed by subtracting the scalar field, which is obtained using the FD method,  from the input field. In comparison with the FFT method shown in Fig.~\ref{fig:Input_Scalar_M1_FFT_single}, the error is larger by more than an order of magnitude.    }
\label{fig:Input_Scalar_M1_FD_single}
\end{figure}

In Fig.~\ref{fig:Input_Scalar_M1_FD_single}, we show the vector part obtained by subtracting the scalar part reconstructed using the FD method. Comparing with Fig.~\ref{fig:Input_Scalar_M1_FFT_single}, we note that the error of the FD method is larger than the FFT method by an order of magnitude. The error is especially large at the boundary, probably because the field does not fall off enough at the boundary, so that imposing periodic boundary condition results in large error.

\begin{figure}[!htb]
\centering
\includegraphics[ width=\linewidth]{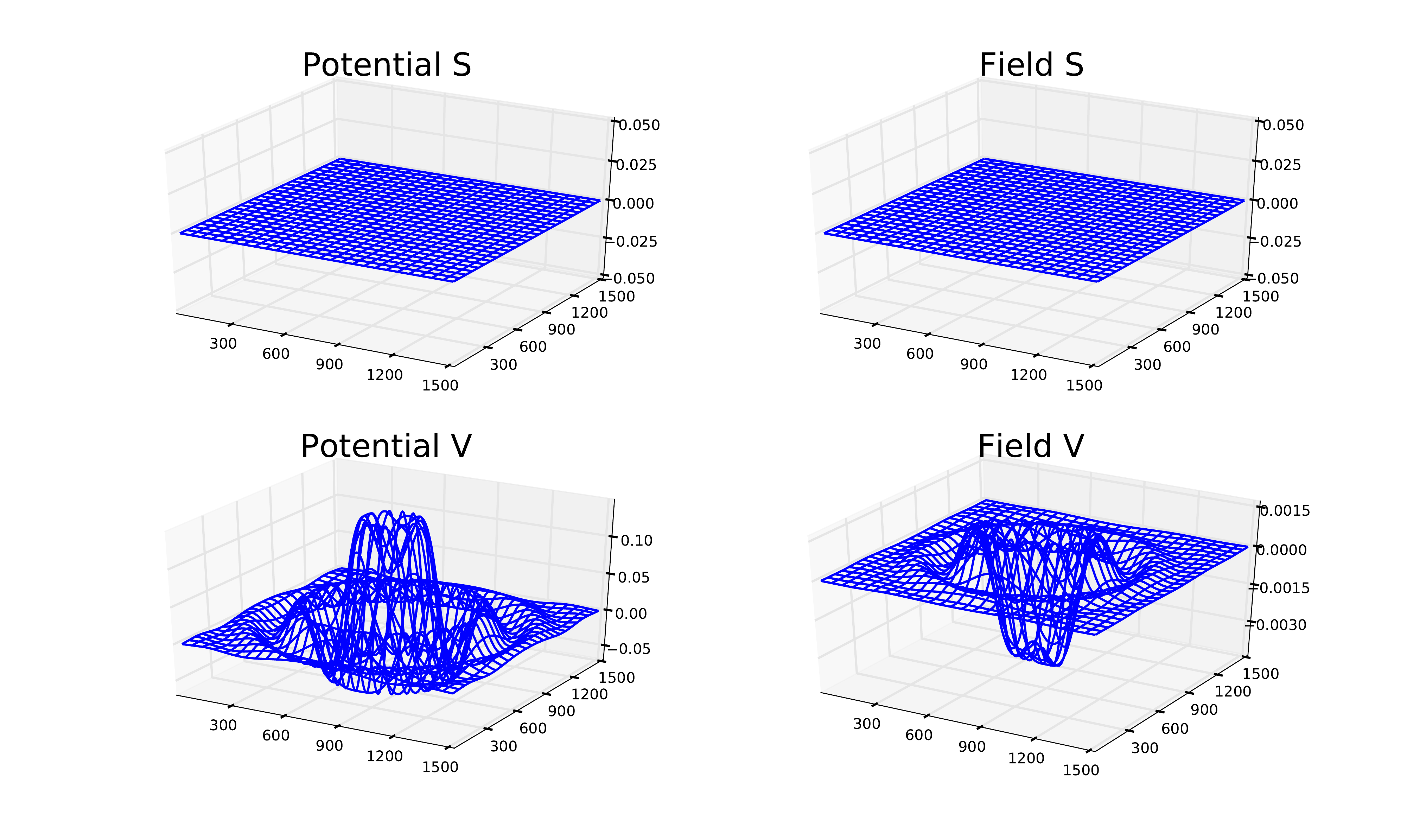}
\caption{  Sections of the scalar part and vector part of the potentials (Eq.~\ref{eq:Test2_Phi} and \ref{eq:Test2_A}) (first column) and its corresponding fields (second column). For the vector field, only the $x$-component is shown.   }
\label{fig:CongfigPlot_M2_input}
\end{figure}

The second example we consider is purely transverse, i.e., it is given by 
\begin{eqnarray}
\label{eq:Test2_Phi}
\Phi( \mb{r} ) &=& 0   \\
\label{eq:Test2_A}
\mb{A} ( \mb{r} )   & =&  a \mathrm{ e} ^{ - r'  / \rho }    \left(
 \begin{array}{c}
 \sin \frac{ r' }{ \sigma }   \\
 \cos \frac{ r' }{ \sigma }   \\
1
 \end{array} \right)
 \end{eqnarray}
The corresponding vector field is given by 
\begin{eqnarray}
\label{eq:VectorTest}
&   &  \mb{V} ( \mb{r} )    =  a \mathrm{ e} ^{ - r'  / \rho }    \\
& \times &     \left(
 \begin{array}{c}
  - \frac{ r_y' }{  \rho r'  }  + \frac{ r_z' }{  \rho r'  } \cos \frac{r' }{ \sigma } +  \frac{ r_z' }{  \sigma  r'  } \sin \frac{r' }{ \sigma }      \\
   \frac{ r_x' }{  \rho r'  }  - \frac{ r_z' }{  \rho r'  } \sin \frac{r' }{ \sigma } +  \frac{ r_z' }{  \sigma  r'  } \cos \frac{r' }{ \sigma }      \\
\Big( - \frac{ r_y' }{  \sigma r'  }  - \frac{ r_x' }{  \rho  r'  }     \Big)  \cos \frac{r' }{ \sigma } + \Big( - \frac{ r_x' }{  \sigma r'  }  + \frac{ r_y' }{  \rho  r'  }     \Big)  \sin \frac{r' }{ \sigma }  
 \end{array} \right) .   \nonumber
 \end{eqnarray}
In Fig.~\ref{fig:CongfigPlot_M2_input} we show the sections of the scalar and vector potentials and the corresponding fields. In this plot the parameters are the same as in the previous test case. In addition, $\sigma=37.5 \MpcOh $.  

\begin{figure}[!htb]
\centering
\includegraphics[ width=\linewidth]{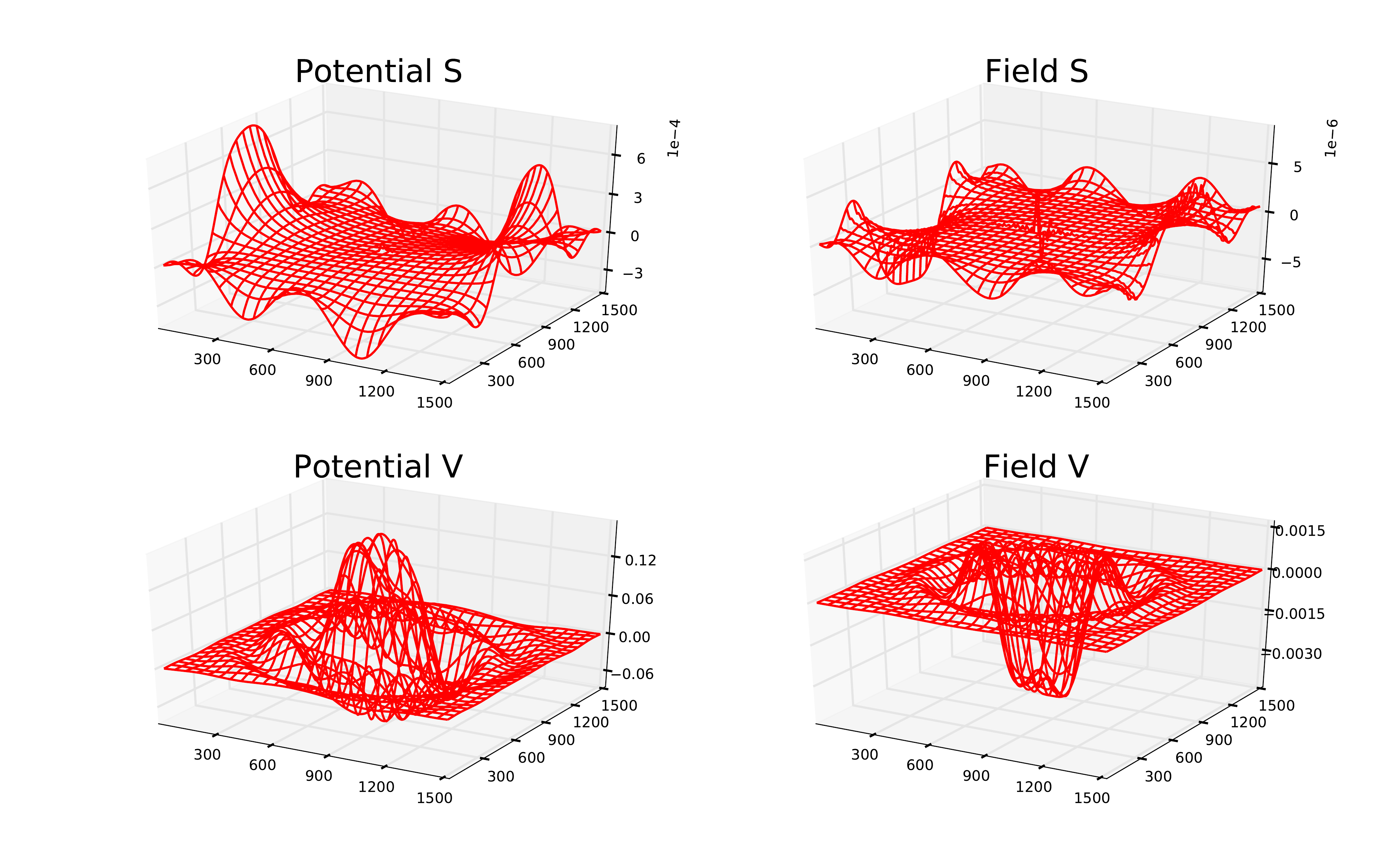}
\caption{  Same as Fig.~\ref{fig:CongfigPlot_M2_input}, but reconstructed using the FFT method.   }
\label{fig:CongfigPlot_M2_reconst_FFT}
\end{figure}

In Fig.~\ref{fig:CongfigPlot_M2_reconst_FFT}, we have shown the potentials and fields reconstructed using the FFT method. The error of the reconstruction, the scalar part of the field,  is about 3 orders of magnitude smaller than the vector part of the field, similar to the previous test case.   

\begin{figure}[!htb]
\centering
\includegraphics[ width=\linewidth]{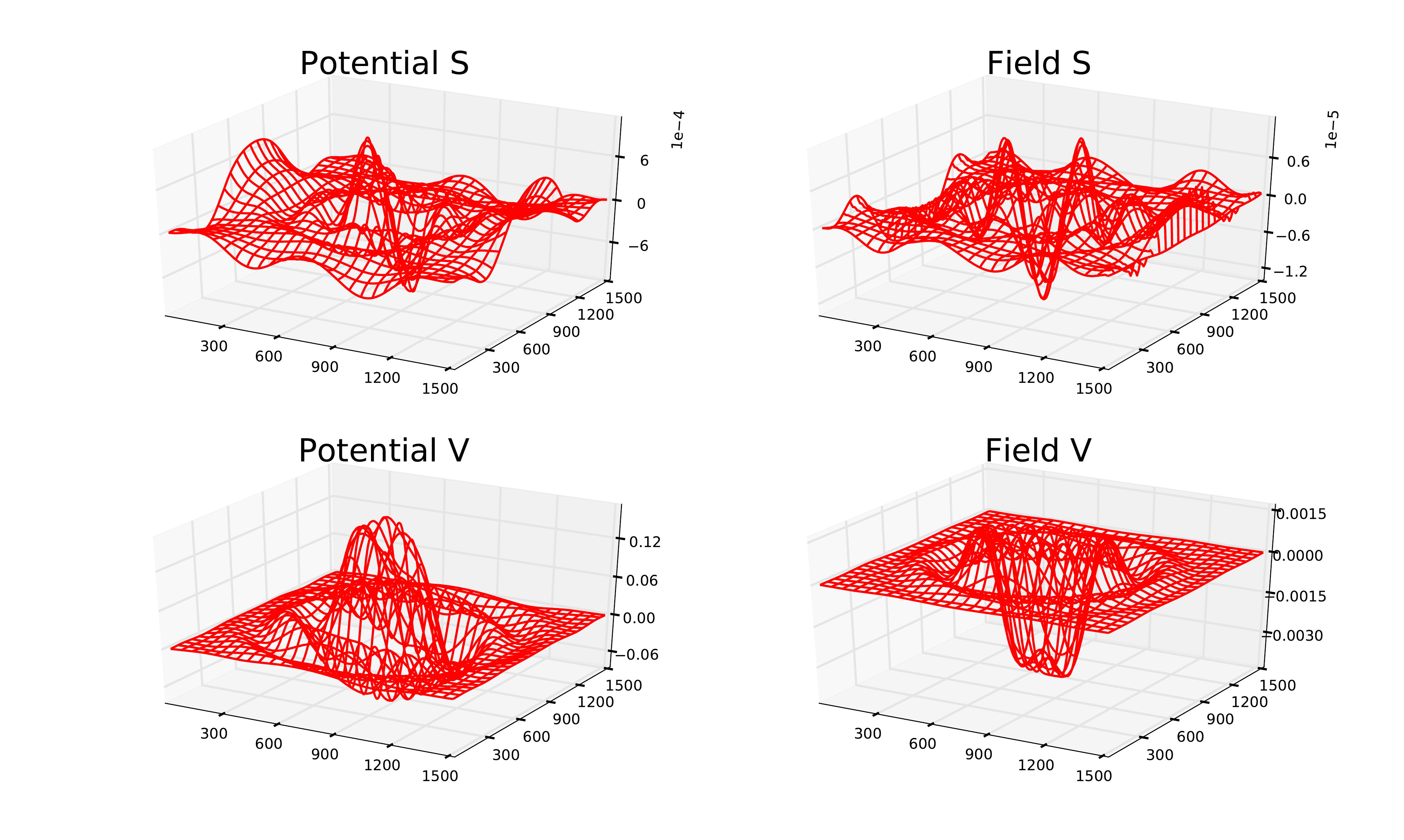}
\caption{  Same as Fig.~\ref{fig:CongfigPlot_M2_input}, but reconstructed using the FD method.   }
\label{fig:CongfigPlot_M2_reconst_FD}
\end{figure}

Finally, in Fig.~\ref{fig:CongfigPlot_M2_reconst_FD}, we show the corresponding results reconstructed using the FD method. Again, the error of the FD method is about twice that of the FFT method. Similar to the scalar case, the error, \textit{i.e.}, the scalar contamination,  is more smooth in the case of FFT method.  


To conclude, the signal in the reconstruction is at least three orders of magnitude higher than the error contamination.  Also we find that the FFT method is more accurate than the FD method by a factor of 2. The vector field obtained by solving the vector potential is very similar to the one obtained by subtracting the scalar component from the input field.


\bibliography{HelmholtzPsi} 

\begin{thebibliography}{47}%
\makeatletter
\providecommand \@ifxundefined [1]{%
 \@ifx{#1\undefined}
}%
\providecommand \@ifnum [1]{%
 \ifnum #1\expandafter \@firstoftwo
 \else \expandafter \@secondoftwo
 \fi
}%
\providecommand \@ifx [1]{%
 \ifx #1\expandafter \@firstoftwo
 \else \expandafter \@secondoftwo
 \fi
}%
\providecommand \natexlab [1]{#1}%
\providecommand \enquote  [1]{``#1''}%
\providecommand \bibnamefont  [1]{#1}%
\providecommand \bibfnamefont [1]{#1}%
\providecommand \citenamefont [1]{#1}%
\providecommand \href@noop [0]{\@secondoftwo}%
\providecommand \href [0]{\begingroup \@sanitize@url \@href}%
\providecommand \@href[1]{\@@startlink{#1}\@@href}%
\providecommand \@@href[1]{\endgroup#1\@@endlink}%
\providecommand \@sanitize@url [0]{\catcode `\\12\catcode `\$12\catcode
  `\&12\catcode `\#12\catcode `\^12\catcode `\_12\catcode `\%12\relax}%
\providecommand \@@startlink[1]{}%
\providecommand \@@endlink[0]{}%
\providecommand \url  [0]{\begingroup\@sanitize@url \@url }%
\providecommand \@url [1]{\endgroup\@href {#1}{\urlprefix }}%
\providecommand \urlprefix  [0]{URL }%
\providecommand \Eprint [0]{\href }%
\providecommand \doibase [0]{http://dx.doi.org/}%
\providecommand \selectlanguage [0]{\@gobble}%
\providecommand \bibinfo  [0]{\@secondoftwo}%
\providecommand \bibfield  [0]{\@secondoftwo}%
\providecommand \translation [1]{[#1]}%
\providecommand \BibitemOpen [0]{}%
\providecommand \bibitemStop [0]{}%
\providecommand \bibitemNoStop [0]{.\EOS\space}%
\providecommand \EOS [0]{\spacefactor3000\relax}%
\providecommand \BibitemShut  [1]{\csname bibitem#1\endcsname}%
\let\auto@bib@innerbib\@empty
\bibitem [{\citenamefont {Scoccimarro}\ and\ \citenamefont
  {Sheth}(2002)}]{ScoccimarroSheth2002}%
  \BibitemOpen
  \bibfield  {author} {\bibinfo {author} {\bibfnamefont {R.}~\bibnamefont
  {Scoccimarro}}\ and\ \bibinfo {author} {\bibfnamefont {R.~K.}\ \bibnamefont
  {Sheth}},\ }\href@noop {} {\bibfield  {journal} {\bibinfo  {journal} {MNRAS}\
  }\textbf {\bibinfo {volume} {329}},\ \bibinfo {pages} {629} (\bibinfo {year}
  {2002})},\ \Eprint {http://arxiv.org/abs/arXiv:astro-ph/0106120}
  {arXiv:astro-ph/0106120} \BibitemShut {NoStop}%
\bibitem [{\citenamefont {Manera}\ \emph {et~al.}(2013)\citenamefont {Manera},
  \citenamefont {Scoccimarro}, \citenamefont {Percival}, \citenamefont
  {Samushia}, \citenamefont {McBride} \emph {et~al.}}]{Maneraetal2013}%
  \BibitemOpen
  \bibfield  {author} {\bibinfo {author} {\bibfnamefont {M.}~\bibnamefont
  {Manera}}, \bibinfo {author} {\bibfnamefont {R.}~\bibnamefont {Scoccimarro}},
  \bibinfo {author} {\bibfnamefont {W.~J.}\ \bibnamefont {Percival}}, \bibinfo
  {author} {\bibfnamefont {L.}~\bibnamefont {Samushia}}, \bibinfo {author}
  {\bibfnamefont {C.~K.}\ \bibnamefont {McBride}},  \emph {et~al.},\
  }\href@noop {} {\bibfield  {journal} {\bibinfo  {journal} {MNRAS}\ }\textbf
  {\bibinfo {volume} {428}},\ \bibinfo {pages} {1036} (\bibinfo {year}
  {2013})},\ \Eprint {http://arxiv.org/abs/arXiv:1203.6609} {arXiv:1203.6609}
  \BibitemShut {NoStop}%
\bibitem [{\citenamefont {Monaco}\ \emph {et~al.}(2002)\citenamefont {Monaco},
  \citenamefont {Theuns},\ and\ \citenamefont {Taffoni}}]{Monacaetal2002}%
  \BibitemOpen
  \bibfield  {author} {\bibinfo {author} {\bibfnamefont {P.}~\bibnamefont
  {Monaco}}, \bibinfo {author} {\bibfnamefont {T.}~\bibnamefont {Theuns}}, \
  and\ \bibinfo {author} {\bibfnamefont {G.}~\bibnamefont {Taffoni}},\
  }\href@noop {} {\bibfield  {journal} {\bibinfo  {journal} {MNRAS}\ }\textbf
  {\bibinfo {volume} {331}},\ \bibinfo {pages} {587} (\bibinfo {year}
  {2002})},\ \Eprint {http://arxiv.org/abs/arXiv:astro-ph/0109323}
  {arXiv:astro-ph/0109323} \BibitemShut {NoStop}%
\bibitem [{\citenamefont {Monaco}\ \emph {et~al.}()\citenamefont {Monaco},
  \citenamefont {Sefusatti}, \citenamefont {Borgani}, \citenamefont {Crocce},
  \citenamefont {Fosalba} \emph {et~al.}}]{Monacoetal2013}%
  \BibitemOpen
  \bibfield  {author} {\bibinfo {author} {\bibfnamefont {P.}~\bibnamefont
  {Monaco}}, \bibinfo {author} {\bibfnamefont {E.}~\bibnamefont {Sefusatti}},
  \bibinfo {author} {\bibfnamefont {S.}~\bibnamefont {Borgani}}, \bibinfo
  {author} {\bibfnamefont {M.}~\bibnamefont {Crocce}}, \bibinfo {author}
  {\bibfnamefont {P.}~\bibnamefont {Fosalba}},  \emph {et~al.},\ }\href@noop {}
  {}\Eprint {http://arxiv.org/abs/arXiv:1305.1505} {arXiv:1305.1505}
  \BibitemShut {NoStop}%
\bibitem [{\citenamefont {Heisenberg}\ \emph {et~al.}(2011)\citenamefont
  {Heisenberg}, \citenamefont {Schaefer},\ and\ \citenamefont
  {Bartelmann}}]{Heisenbergetal2011}%
  \BibitemOpen
  \bibfield  {author} {\bibinfo {author} {\bibfnamefont {L.}~\bibnamefont
  {Heisenberg}}, \bibinfo {author} {\bibfnamefont {B.~M.}\ \bibnamefont
  {Schaefer}}, \ and\ \bibinfo {author} {\bibfnamefont {M.}~\bibnamefont
  {Bartelmann}},\ }\href@noop {} {\bibfield  {journal} {\bibinfo  {journal}
  {MNRAS}\ }\textbf {\bibinfo {volume} {416}},\ \bibinfo {pages} {3057}
  (\bibinfo {year} {2011})},\ \Eprint {http://arxiv.org/abs/arXiv:1011.1559}
  {arXiv:1011.1559} \BibitemShut {NoStop}%
\bibitem [{\citenamefont {Tassev}\ \emph {et~al.}()\citenamefont {Tassev},
  \citenamefont {Zaldarriaga},\ and\ \citenamefont
  {Eisenstein}}]{Tsssevetal2013}%
  \BibitemOpen
  \bibfield  {author} {\bibinfo {author} {\bibfnamefont {S.}~\bibnamefont
  {Tassev}}, \bibinfo {author} {\bibfnamefont {M.}~\bibnamefont {Zaldarriaga}},
  \ and\ \bibinfo {author} {\bibfnamefont {D.}~\bibnamefont {Eisenstein}},\
  }\href@noop {} {}\Eprint {http://arxiv.org/abs/arXiv:1301.0322}
  {arXiv:1301.0322} \BibitemShut {NoStop}%
\bibitem [{\citenamefont {Zel'dovich}(1970)}]{Zeldovich1970}%
  \BibitemOpen
  \bibfield  {author} {\bibinfo {author} {\bibfnamefont {Y.~B.}\ \bibnamefont
  {Zel'dovich}},\ }\href@noop {} {\bibfield  {journal} {\bibinfo  {journal}
  {A\&A}\ }\textbf {\bibinfo {volume} {5}},\ \bibinfo {pages} {84} (\bibinfo
  {year} {1970})}\BibitemShut {NoStop}%
\bibitem [{\citenamefont {Buchert}(1994)}]{Buchert1994}%
  \BibitemOpen
  \bibfield  {author} {\bibinfo {author} {\bibfnamefont {T.}~\bibnamefont
  {Buchert}},\ }\href@noop {} {\bibfield  {journal} {\bibinfo  {journal}
  {MNRAS}\ }\textbf {\bibinfo {volume} {267}},\ \bibinfo {pages} {811}
  (\bibinfo {year} {1994})}\BibitemShut {NoStop}%
\bibitem [{\citenamefont {Catelan}(1995)}]{Catelan95}%
  \BibitemOpen
  \bibfield  {author} {\bibinfo {author} {\bibfnamefont {P.}~\bibnamefont
  {Catelan}},\ }\href@noop {} {\bibfield  {journal} {\bibinfo  {journal}
  {MNRAS}\ }\textbf {\bibinfo {volume} {276}},\ \bibinfo {pages} {115}
  (\bibinfo {year} {1995})}\BibitemShut {NoStop}%
\bibitem [{\citenamefont {Catelan}\ and\ \citenamefont
  {Theuns}(1996)}]{CatelanTheuns96}%
  \BibitemOpen
  \bibfield  {author} {\bibinfo {author} {\bibfnamefont {P.}~\bibnamefont
  {Catelan}}\ and\ \bibinfo {author} {\bibfnamefont {T.}~\bibnamefont
  {Theuns}},\ }\href@noop {} {\bibfield  {journal} {\bibinfo  {journal}
  {MNRAS}\ }\textbf {\bibinfo {volume} {282}},\ \bibinfo {pages} {455}
  (\bibinfo {year} {1996})}\BibitemShut {NoStop}%
\bibitem [{\citenamefont {Bouchet}\ \emph {et~al.}(1995)\citenamefont
  {Bouchet}, \citenamefont {Colombi}, \citenamefont {Hivon},\ and\
  \citenamefont {Juszkiewicz}}]{Bouchetetal1995}%
  \BibitemOpen
  \bibfield  {author} {\bibinfo {author} {\bibfnamefont {F.~R.}\ \bibnamefont
  {Bouchet}}, \bibinfo {author} {\bibfnamefont {S.}~\bibnamefont {Colombi}},
  \bibinfo {author} {\bibfnamefont {E.}~\bibnamefont {Hivon}}, \ and\ \bibinfo
  {author} {\bibfnamefont {R.}~\bibnamefont {Juszkiewicz}},\ }\href@noop {}
  {\bibfield  {journal} {\bibinfo  {journal} {A\&A}\ }\textbf {\bibinfo
  {volume} {296}},\ \bibinfo {pages} {575} (\bibinfo {year} {1995})},\ \Eprint
  {http://arxiv.org/abs/arXiv:astro-ph/9406013} {arXiv:astro-ph/9406013}
  \BibitemShut {NoStop}%
\bibitem [{\citenamefont {Rampf}\ and\ \citenamefont
  {Buchert}(2012)}]{RampfBuchert12}%
  \BibitemOpen
  \bibfield  {author} {\bibinfo {author} {\bibfnamefont {C.}~\bibnamefont
  {Rampf}}\ and\ \bibinfo {author} {\bibfnamefont {T.}~\bibnamefont
  {Buchert}},\ }\href@noop {} {\bibfield  {journal} {\bibinfo  {journal}
  {JCAP}\ }\textbf {\bibinfo {volume} {06}},\ \bibinfo {pages} {021} (\bibinfo
  {year} {2012})},\ \Eprint {http://arxiv.org/abs/arXiv:1203.4260}
  {arXiv:1203.4260} \BibitemShut {NoStop}%
\bibitem [{\citenamefont {Rampf}(2012)}]{Rampf2012}%
  \BibitemOpen
  \bibfield  {author} {\bibinfo {author} {\bibfnamefont {C.}~\bibnamefont
  {Rampf}},\ }\href@noop {} {\bibfield  {journal} {\bibinfo  {journal} {JCAP}\
  }\textbf {\bibinfo {volume} {12}},\ \bibinfo {pages} {4} (\bibinfo {year}
  {2012})},\ \Eprint {http://arxiv.org/abs/arXiv:1205.5274} {arXiv:1205.5274}
  \BibitemShut {NoStop}%
\bibitem [{\citenamefont {Scoccimarro}(1998)}]{Scoccimarro98}%
  \BibitemOpen
  \bibfield  {author} {\bibinfo {author} {\bibfnamefont {R.}~\bibnamefont
  {Scoccimarro}},\ }\href@noop {} {\bibfield  {journal} {\bibinfo  {journal}
  {MNRAS}\ }\textbf {\bibinfo {volume} {299}},\ \bibinfo {pages} {1097}
  (\bibinfo {year} {1998})},\ \Eprint
  {http://arxiv.org/abs/arXiv:astro-ph/9711187} {arXiv:astro-ph/9711187}
  \BibitemShut {NoStop}%
\bibitem [{\citenamefont {Crocce}\ \emph {et~al.}(2006)\citenamefont {Crocce},
  \citenamefont {Pueblas},\ and\ \citenamefont
  {Scoccimarro}}]{CroccePeublasetal2006}%
  \BibitemOpen
  \bibfield  {author} {\bibinfo {author} {\bibfnamefont {M.}~\bibnamefont
  {Crocce}}, \bibinfo {author} {\bibfnamefont {S.}~\bibnamefont {Pueblas}}, \
  and\ \bibinfo {author} {\bibfnamefont {R.}~\bibnamefont {Scoccimarro}},\
  }\href@noop {} {\bibfield  {journal} {\bibinfo  {journal} {MNRAS}\ }\textbf
  {\bibinfo {volume} {373}},\ \bibinfo {pages} {369} (\bibinfo {year}
  {2006})},\ \Eprint {http://arxiv.org/abs/arXiv:astro-ph/0606505}
  {arXiv:astro-ph/0606505} \BibitemShut {NoStop}%
\bibitem [{\citenamefont {Matsubara}(2008{\natexlab{a}})}]{Matsubara08a}%
  \BibitemOpen
  \bibfield  {author} {\bibinfo {author} {\bibfnamefont {T.}~\bibnamefont
  {Matsubara}},\ }\href@noop {} {\bibfield  {journal} {\bibinfo  {journal}
  {Phys. Rev. D}\ }\textbf {\bibinfo {volume} {77}},\ \bibinfo {pages} {063530}
  (\bibinfo {year} {2008}{\natexlab{a}})},\ \Eprint
  {http://arxiv.org/abs/arXiv:0711.2521} {arXiv:0711.2521} \BibitemShut
  {NoStop}%
\bibitem [{\citenamefont {Matsubara}(2008{\natexlab{b}})}]{Matsubara08b}%
  \BibitemOpen
  \bibfield  {author} {\bibinfo {author} {\bibfnamefont {T.}~\bibnamefont
  {Matsubara}},\ }\href@noop {} {\bibfield  {journal} {\bibinfo  {journal}
  {Phys. Rev. D}\ }\textbf {\bibinfo {volume} {78}},\ \bibinfo {pages} {083519}
  (\bibinfo {year} {2008}{\natexlab{b}})},\ \Eprint
  {http://arxiv.org/abs/arXiv:0807.1733} {arXiv:0807.1733} \BibitemShut
  {NoStop}%
\bibitem [{\citenamefont {Matsubara}(2011)}]{Matsubara11}%
  \BibitemOpen
  \bibfield  {author} {\bibinfo {author} {\bibfnamefont {T.}~\bibnamefont
  {Matsubara}},\ }\href@noop {} {\bibfield  {journal} {\bibinfo  {journal}
  {Phys. Rev. D}\ }\textbf {\bibinfo {volume} {83}},\ \bibinfo {pages} {083518}
  (\bibinfo {year} {2011})},\ \Eprint {http://arxiv.org/abs/arXiv:1102.4619}
  {arXiv:1102.4619} \BibitemShut {NoStop}%
\bibitem [{\citenamefont {Bernardeau}\ \emph {et~al.}(2002)\citenamefont
  {Bernardeau}, \citenamefont {Colombi}, \citenamefont {Gazta{\~n}aga},\ and\
  \citenamefont {Scoccimarro}}]{PTreview}%
  \BibitemOpen
  \bibfield  {author} {\bibinfo {author} {\bibfnamefont {F.}~\bibnamefont
  {Bernardeau}}, \bibinfo {author} {\bibfnamefont {S.}~\bibnamefont {Colombi}},
  \bibinfo {author} {\bibfnamefont {E.}~\bibnamefont {Gazta{\~n}aga}}, \ and\
  \bibinfo {author} {\bibfnamefont {R.}~\bibnamefont {Scoccimarro}},\
  }\href@noop {} {\bibfield  {journal} {\bibinfo  {journal} {Phys. Rep.}\
  }\textbf {\bibinfo {volume} {367}},\ \bibinfo {pages} {1} (\bibinfo {year}
  {2002})},\ \Eprint {http://arxiv.org/abs/arXiv:astro-ph/0112551}
  {arXiv:astro-ph/0112551} \BibitemShut {NoStop}%
\bibitem [{\citenamefont {Pueblas}\ and\ \citenamefont
  {Scoccimarro}(2009)}]{PueblasScoccimarro2009}%
  \BibitemOpen
  \bibfield  {author} {\bibinfo {author} {\bibfnamefont {S.}~\bibnamefont
  {Pueblas}}\ and\ \bibinfo {author} {\bibfnamefont {R.}~\bibnamefont
  {Scoccimarro}},\ }\href@noop {} {\bibfield  {journal} {\bibinfo  {journal}
  {Phys. Rev. D}\ }\textbf {\bibinfo {volume} {80}},\ \bibinfo {pages} {043504}
  (\bibinfo {year} {2009})}\BibitemShut {NoStop}%
\bibitem [{\citenamefont {Shandarin}\ and\ \citenamefont
  {Zel'dovich}(1989)}]{ShandarinZeldovich1989}%
  \BibitemOpen
  \bibfield  {author} {\bibinfo {author} {\bibfnamefont {S.~F.}\ \bibnamefont
  {Shandarin}}\ and\ \bibinfo {author} {\bibfnamefont {Y.~B.}\ \bibnamefont
  {Zel'dovich}},\ }\href@noop {} {\bibfield  {journal} {\bibinfo  {journal}
  {Rev. Mod. Phys.}\ }\textbf {\bibinfo {volume} {61}},\ \bibinfo {pages} {185}
  (\bibinfo {year} {1989})}\BibitemShut {NoStop}%
\bibitem [{\citenamefont {Melott}\ \emph {et~al.}(1994)\citenamefont {Melott},
  \citenamefont {Pellman},\ and\ \citenamefont
  {Shandarin}}]{MelottPellmanetal1994}%
  \BibitemOpen
  \bibfield  {author} {\bibinfo {author} {\bibfnamefont {A.~L.}\ \bibnamefont
  {Melott}}, \bibinfo {author} {\bibfnamefont {T.~F.}\ \bibnamefont {Pellman}},
  \ and\ \bibinfo {author} {\bibfnamefont {S.~F.}\ \bibnamefont {Shandarin}},\
  }\href@noop {} {\bibfield  {journal} {\bibinfo  {journal} {MNRAS}\ }\textbf
  {\bibinfo {volume} {269}},\ \bibinfo {pages} {626} (\bibinfo {year}
  {1994})},\ \Eprint {http://arxiv.org/abs/arXiv:9312044} {arXiv:9312044}
  \BibitemShut {NoStop}%
\bibitem [{\citenamefont {Kitaura}\ and\ \citenamefont
  {Steffen}(2012)}]{KitauraSreffen2012}%
  \BibitemOpen
  \bibfield  {author} {\bibinfo {author} {\bibfnamefont {F.-S.}\ \bibnamefont
  {Kitaura}}\ and\ \bibinfo {author} {\bibfnamefont {H.}~\bibnamefont
  {Steffen}},\ }\href@noop {} {} (\bibinfo {year} {2012}),\ \Eprint
  {http://arxiv.org/abs/arXiv:1212.3514} {arXiv:1212.3514} \BibitemShut
  {NoStop}%
\bibitem [{\citenamefont {Leclercq}\ \emph {et~al.}(2012)\citenamefont
  {Leclercq}, \citenamefont {Jasche}, \citenamefont {Gil-Marin},\ and\
  \citenamefont {Wandelt}}]{Leclerceetal2013}%
  \BibitemOpen
  \bibfield  {author} {\bibinfo {author} {\bibfnamefont {F.}~\bibnamefont
  {Leclercq}}, \bibinfo {author} {\bibfnamefont {J.}~\bibnamefont {Jasche}},
  \bibinfo {author} {\bibfnamefont {H.}~\bibnamefont {Gil-Marin}}, \ and\
  \bibinfo {author} {\bibfnamefont {B.}~\bibnamefont {Wandelt}},\ }\href@noop
  {} {} (\bibinfo {year} {2012}),\ \Eprint
  {http://arxiv.org/abs/arXiv:1305.4642} {arXiv:1305.4642} \BibitemShut
  {NoStop}%
\bibitem [{\citenamefont {Gurbatov}\ \emph {et~al.}(1989)\citenamefont
  {Gurbatov}, \citenamefont {Saichev},\ and\ \citenamefont
  {Shandarin}}]{Gurbatovetal1989}%
  \BibitemOpen
  \bibfield  {author} {\bibinfo {author} {\bibfnamefont {S.~N.}\ \bibnamefont
  {Gurbatov}}, \bibinfo {author} {\bibfnamefont {A.~I.}\ \bibnamefont
  {Saichev}}, \ and\ \bibinfo {author} {\bibfnamefont {S.~F.}\ \bibnamefont
  {Shandarin}},\ }\href@noop {} {\bibfield  {journal} {\bibinfo  {journal}
  {MNRAS}\ }\textbf {\bibinfo {volume} {236}},\ \bibinfo {pages} {385}
  (\bibinfo {year} {1989})}\BibitemShut {NoStop}%
\bibitem [{\citenamefont {Vergassola}\ \emph {et~al.}(1994)\citenamefont
  {Vergassola}, \citenamefont {Dubrulle}, \citenamefont {Frisch},\ and\
  \citenamefont {Noullez}}]{Vergassolaetal1994}%
  \BibitemOpen
  \bibfield  {author} {\bibinfo {author} {\bibfnamefont {M.}~\bibnamefont
  {Vergassola}}, \bibinfo {author} {\bibfnamefont {B.}~\bibnamefont
  {Dubrulle}}, \bibinfo {author} {\bibfnamefont {U.}~\bibnamefont {Frisch}}, \
  and\ \bibinfo {author} {\bibfnamefont {A.}~\bibnamefont {Noullez}},\
  }\href@noop {} {\bibfield  {journal} {\bibinfo  {journal} {A\&A}\ }\textbf
  {\bibinfo {volume} {289}},\ \bibinfo {pages} {325} (\bibinfo {year}
  {1994})}\BibitemShut {NoStop}%
\bibitem [{\citenamefont {Valageas}(2011)}]{Valageas2011}%
  \BibitemOpen
  \bibfield  {author} {\bibinfo {author} {\bibfnamefont {P.}~\bibnamefont
  {Valageas}},\ }\href@noop {} {\bibfield  {journal} {\bibinfo  {journal}
  {A\&A}\ }\textbf {\bibinfo {volume} {526}},\ \bibinfo {pages} {67} (\bibinfo
  {year} {2011})}\BibitemShut {NoStop}%
\bibitem [{\citenamefont {Weinberg}\ and\ \citenamefont
  {Gunn}(1990)}]{WeinbergGunn}%
  \BibitemOpen
  \bibfield  {author} {\bibinfo {author} {\bibfnamefont {D.~H.}\ \bibnamefont
  {Weinberg}}\ and\ \bibinfo {author} {\bibfnamefont {J.~E.}\ \bibnamefont
  {Gunn}},\ }\href@noop {} {\bibfield  {journal} {\bibinfo  {journal} {MNRAS}\
  }\textbf {\bibinfo {volume} {247}},\ \bibinfo {pages} {260} (\bibinfo {year}
  {1990})}\BibitemShut {NoStop}%
\bibitem [{\citenamefont {Jackson}(1998)}]{Jackson}%
  \BibitemOpen
  \bibfield  {author} {\bibinfo {author} {\bibfnamefont {J.~D.}\ \bibnamefont
  {Jackson}},\ }\href@noop {} {\emph {\bibinfo {title} {Classical
  Electrodynamics Third Edition}}}\ (\bibinfo  {publisher} {Wiley},\ \bibinfo
  {year} {1998})\BibitemShut {NoStop}%
\bibitem [{\citenamefont {Bertschinger}()}]{Bertschinger1995}%
  \BibitemOpen
  \bibfield  {author} {\bibinfo {author} {\bibfnamefont {E.}~\bibnamefont
  {Bertschinger}},\ }\href@noop {} {}\Eprint
  {http://arxiv.org/abs/arXiv:9503125} {arXiv:9503125} \BibitemShut {NoStop}%
\bibitem [{\citenamefont {Zhang}\ \emph {et~al.}(2013)\citenamefont {Zhang},
  \citenamefont {Pan},\ and\ \citenamefont {Zheng}}]{Zhangetal2013}%
  \BibitemOpen
  \bibfield  {author} {\bibinfo {author} {\bibfnamefont {P.}~\bibnamefont
  {Zhang}}, \bibinfo {author} {\bibfnamefont {J.}~\bibnamefont {Pan}}, \ and\
  \bibinfo {author} {\bibfnamefont {Y.}~\bibnamefont {Zheng}},\ }\href@noop {}
  {\bibfield  {journal} {\bibinfo  {journal} {Phys. Rev. D}\ }\textbf {\bibinfo
  {volume} {87}},\ \bibinfo {pages} {063526} (\bibinfo {year} {2013})},\
  \Eprint {http://arxiv.org/abs/arXiv:1207.2722} {arXiv:1207.2722} \BibitemShut
  {NoStop}%
\bibitem [{\citenamefont {Zheng}\ \emph {et~al.}()\citenamefont {Zheng},
  \citenamefont {Zhang}, \citenamefont {Jing}, \citenamefont {Lin},\ and\
  \citenamefont {Pan}}]{Zhengetal2013}%
  \BibitemOpen
  \bibfield  {author} {\bibinfo {author} {\bibfnamefont {Y.}~\bibnamefont
  {Zheng}}, \bibinfo {author} {\bibfnamefont {P.}~\bibnamefont {Zhang}},
  \bibinfo {author} {\bibfnamefont {Y.}~\bibnamefont {Jing}}, \bibinfo {author}
  {\bibfnamefont {W.}~\bibnamefont {Lin}}, \ and\ \bibinfo {author}
  {\bibfnamefont {J.}~\bibnamefont {Pan}},\ }\href@noop {} {}\Eprint
  {http://arxiv.org/abs/arXiv:1308.0886} {arXiv:1308.0886} \BibitemShut
  {NoStop}%
\bibitem [{\citenamefont {Buchert}\ \emph {et~al.}(1994)\citenamefont
  {Buchert}, \citenamefont {Melott},\ and\ \citenamefont
  {Weiss}}]{Buchertetal1994}%
  \BibitemOpen
  \bibfield  {author} {\bibinfo {author} {\bibfnamefont {T.}~\bibnamefont
  {Buchert}}, \bibinfo {author} {\bibfnamefont {A.~L.}\ \bibnamefont {Melott}},
  \ and\ \bibinfo {author} {\bibfnamefont {A.~G.}\ \bibnamefont {Weiss}},\
  }\href@noop {} {\bibfield  {journal} {\bibinfo  {journal} {A\&A}\ }\textbf
  {\bibinfo {volume} {288}},\ \bibinfo {pages} {349} (\bibinfo {year}
  {1994})}\BibitemShut {NoStop}%
\bibitem [{\citenamefont {Bernardeau}\ and\ \citenamefont {van~de
  Weygaert}(1996)}]{BernardeauWeygaert1996}%
  \BibitemOpen
  \bibfield  {author} {\bibinfo {author} {\bibfnamefont {F.}~\bibnamefont
  {Bernardeau}}\ and\ \bibinfo {author} {\bibfnamefont {R.}~\bibnamefont
  {van~de Weygaert}},\ }\href@noop {} {\bibfield  {journal} {\bibinfo
  {journal} {MNRAS}\ }\textbf {\bibinfo {volume} {279}},\ \bibinfo {pages}
  {693} (\bibinfo {year} {1996})}\BibitemShut {NoStop}%
\bibitem [{\citenamefont {van~de Weygaert}\ and\ \citenamefont
  {Schaap}()}]{WeygaertSchaap}%
  \BibitemOpen
  \bibfield  {author} {\bibinfo {author} {\bibfnamefont {R.}~\bibnamefont
  {van~de Weygaert}}\ and\ \bibinfo {author} {\bibfnamefont {W.}~\bibnamefont
  {Schaap}},\ }\href@noop {} {}\Eprint {http://arxiv.org/abs/arXiv:0708.1441}
  {arXiv:0708.1441} \BibitemShut {NoStop}%
\bibitem [{\citenamefont {Komatsu}\ \emph {et~al.}(2011)\citenamefont
  {Komatsu}, \citenamefont {Smith}, \citenamefont {Dunkley} \emph
  {et~al.}}]{WMAP7}%
  \BibitemOpen
  \bibfield  {author} {\bibinfo {author} {\bibfnamefont {E.}~\bibnamefont
  {Komatsu}}, \bibinfo {author} {\bibfnamefont {K.~M.}\ \bibnamefont {Smith}},
  \bibinfo {author} {\bibfnamefont {J.}~\bibnamefont {Dunkley}},  \emph
  {et~al.},\ }\href@noop {} {\bibfield  {journal} {\bibinfo  {journal} {ApJ.
  Suppl.}\ }\textbf {\bibinfo {volume} {192}},\ \bibinfo {pages} {18} (\bibinfo
  {year} {2011})},\ \Eprint {http://arxiv.org/abs/arXiv:1001.4538}
  {arXiv:1001.4538} \BibitemShut {NoStop}%
\bibitem [{\citenamefont {Lewis}\ \emph {et~al.}(2000)\citenamefont {Lewis},
  \citenamefont {Challinor},\ and\ \citenamefont {Lasenby}}]{CAMB}%
  \BibitemOpen
  \bibfield  {author} {\bibinfo {author} {\bibfnamefont {A.}~\bibnamefont
  {Lewis}}, \bibinfo {author} {\bibfnamefont {A.}~\bibnamefont {Challinor}}, \
  and\ \bibinfo {author} {\bibfnamefont {A.}~\bibnamefont {Lasenby}},\
  }\href@noop {} {\bibfield  {journal} {\bibinfo  {journal} {ApJ.}\ }\textbf
  {\bibinfo {volume} {538}},\ \bibinfo {pages} {473} (\bibinfo {year}
  {2000})},\ \Eprint {http://arxiv.org/abs/arXiv:astro-ph/9911177}
  {arXiv:astro-ph/9911177} \BibitemShut {NoStop}%
\bibitem [{\citenamefont {Springel}(2005)}]{Gadget2}%
  \BibitemOpen
  \bibfield  {author} {\bibinfo {author} {\bibfnamefont {V.}~\bibnamefont
  {Springel}},\ }\href@noop {} {\bibfield  {journal} {\bibinfo  {journal}
  {MNRAS}\ }\textbf {\bibinfo {volume} {364}},\ \bibinfo {pages} {1105}
  (\bibinfo {year} {2005})},\ \Eprint
  {http://arxiv.org/abs/arXiv:astro-ph/0505010} {arXiv:astro-ph/0505010}
  \BibitemShut {NoStop}%
\bibitem [{\citenamefont {Biagetti}\ \emph {et~al.}()\citenamefont {Biagetti},
  \citenamefont {Chan}, \citenamefont {Desjacques},\ and\ \citenamefont
  {Paranjape}}]{Biagetietal2013}%
  \BibitemOpen
  \bibfield  {author} {\bibinfo {author} {\bibfnamefont {M.}~\bibnamefont
  {Biagetti}}, \bibinfo {author} {\bibfnamefont {K.~C.}\ \bibnamefont {Chan}},
  \bibinfo {author} {\bibfnamefont {V.}~\bibnamefont {Desjacques}}, \ and\
  \bibinfo {author} {\bibfnamefont {A.}~\bibnamefont {Paranjape}},\ }\href@noop
  {} {}\Eprint {http://arxiv.org/abs/arXiv:1310.1401} {arXiv:1310.1401}
  \BibitemShut {NoStop}%
\bibitem [{\citenamefont {Pichon}\ and\ \citenamefont
  {Bernardeau}(1999)}]{PichonBernardeau}%
  \BibitemOpen
  \bibfield  {author} {\bibinfo {author} {\bibfnamefont {C.}~\bibnamefont
  {Pichon}}\ and\ \bibinfo {author} {\bibfnamefont {F.}~\bibnamefont
  {Bernardeau}},\ }\href@noop {} {\bibfield  {journal} {\bibinfo  {journal}
  {A\&A}\ }\textbf {\bibinfo {volume} {343}},\ \bibinfo {pages} {663} (\bibinfo
  {year} {1999})},\ \Eprint {http://arxiv.org/abs/arXiv:astro-ph/9902142}
  {arXiv:astro-ph/9902142} \BibitemShut {NoStop}%
\bibitem [{\citenamefont {Berbardeau}(1994)}]{Berbardeau1994}%
  \BibitemOpen
  \bibfield  {author} {\bibinfo {author} {\bibfnamefont {F.}~\bibnamefont
  {Berbardeau}},\ }\href@noop {} {\bibfield  {journal} {\bibinfo  {journal}
  {ApJ}\ }\textbf {\bibinfo {volume} {427}},\ \bibinfo {pages} {51} (\bibinfo
  {year} {1994})},\ \Eprint {http://arxiv.org/abs/arXiv:astro-ph/9311066}
  {arXiv:astro-ph/9311066} \BibitemShut {NoStop}%
\bibitem [{\citenamefont {Mohayaee}\ \emph {et~al.}(2006)\citenamefont
  {Mohayaee}, \citenamefont {Mathis}, \citenamefont {Colombi},\ and\
  \citenamefont {Silk}}]{Mohayaeeetal2006}%
  \BibitemOpen
  \bibfield  {author} {\bibinfo {author} {\bibfnamefont {R.}~\bibnamefont
  {Mohayaee}}, \bibinfo {author} {\bibfnamefont {H.}~\bibnamefont {Mathis}},
  \bibinfo {author} {\bibfnamefont {S.}~\bibnamefont {Colombi}}, \ and\
  \bibinfo {author} {\bibfnamefont {J.}~\bibnamefont {Silk}},\ }\href@noop {}
  {\bibfield  {journal} {\bibinfo  {journal} {MNRAS}\ }\textbf {\bibinfo
  {volume} {365}},\ \bibinfo {pages} {939} (\bibinfo {year} {2006})},\ \Eprint
  {http://arxiv.org/abs/arXiv:astro-ph/0501217} {arXiv:astro-ph/0501217}
  \BibitemShut {NoStop}%
\bibitem [{\citenamefont {Neyrinck}(2012)}]{Neyrinck2012}%
  \BibitemOpen
  \bibfield  {author} {\bibinfo {author} {\bibfnamefont {M.}~\bibnamefont
  {Neyrinck}},\ }\href@noop {} {\bibfield  {journal} {\bibinfo  {journal}
  {MNRAS}\ }\textbf {\bibinfo {volume} {428}},\ \bibinfo {pages} {141}
  (\bibinfo {year} {2012})},\ \Eprint {http://arxiv.org/abs/arXiv:1204.1326}
  {arXiv:1204.1326} \BibitemShut {NoStop}%
\bibitem [{\citenamefont {Lee}\ and\ \citenamefont
  {Shandarin}(1998)}]{LeeShandarin1998}%
  \BibitemOpen
  \bibfield  {author} {\bibinfo {author} {\bibfnamefont {J.}~\bibnamefont
  {Lee}}\ and\ \bibinfo {author} {\bibfnamefont {S.}~\bibnamefont
  {Shandarin}},\ }\href@noop {} {\bibfield  {journal} {\bibinfo  {journal}
  {ApJ}\ }\textbf {\bibinfo {volume} {500}},\ \bibinfo {pages} {14} (\bibinfo
  {year} {1998})}\BibitemShut {NoStop}%
\bibitem [{\citenamefont {Cautun}\ \emph {et~al.}(2013)\citenamefont {Cautun},
  \citenamefont {van~de Weygaert},\ and\ \citenamefont
  {Jones}}]{Cautunetal2013}%
  \BibitemOpen
  \bibfield  {author} {\bibinfo {author} {\bibfnamefont {M.}~\bibnamefont
  {Cautun}}, \bibinfo {author} {\bibfnamefont {R.}~\bibnamefont {van~de
  Weygaert}}, \ and\ \bibinfo {author} {\bibfnamefont {B.}~\bibnamefont
  {Jones}},\ }\href@noop {} {\bibfield  {journal} {\bibinfo  {journal} {MNRAS}\
  }\textbf {\bibinfo {volume} {429}},\ \bibinfo {pages} {1286} (\bibinfo {year}
  {2013})},\ \Eprint {http://arxiv.org/abs/arXiv:1209.2043} {arXiv:1209.2043}
  \BibitemShut {NoStop}%
\bibitem [{\citenamefont {Tempel}\ \emph {et~al.}()\citenamefont {Tempel},
  \citenamefont {Stoica}, \citenamefont {Saar}, \citenamefont {Martinez} \emph
  {et~al.}}]{TempelStoicaetal2013}%
  \BibitemOpen
  \bibfield  {author} {\bibinfo {author} {\bibfnamefont {E.}~\bibnamefont
  {Tempel}}, \bibinfo {author} {\bibfnamefont {R.~S.}\ \bibnamefont {Stoica}},
  \bibinfo {author} {\bibfnamefont {E.}~\bibnamefont {Saar}}, \bibinfo {author}
  {\bibfnamefont {V.~J.}\ \bibnamefont {Martinez}},  \emph {et~al.},\
  }\href@noop {} {}\Eprint {http://arxiv.org/abs/arXiv:1308.2533}
  {arXiv:1308.2533} \BibitemShut {NoStop}%
\bibitem [{\citenamefont {Doroshkevich}(1970)}]{Doroshkevich1970}%
  \BibitemOpen
  \bibfield  {author} {\bibinfo {author} {\bibfnamefont {A.~G.}\ \bibnamefont
  {Doroshkevich}},\ }\href@noop {} {\bibfield  {journal} {\bibinfo  {journal}
  {Astrofizika}\ }\textbf {\bibinfo {volume} {6}},\ \bibinfo {pages} {581}
  (\bibinfo {year} {1970})}\BibitemShut {NoStop}%
\end{thebibliography}%

\end{document}